\documentclass[11pt]{article}
\usepackage{amsfonts}
 \usepackage{amssymb}
 \def\bsh{\backslash}

 \newfont{\bbbold}{msbm10}
 \def\com{\mbox{\bbbold C}}
 \def\real{\mbox{\bbbold R}}

 \def\bbC{\mbox{\bbbold C}}

 \def\bbU{\mbox{\bbbold U}}

 \def\cE{{\cal E}}

 \def\cI{{\cal I}}

 \def\cN{{\cal N}}
 \def\cO{{\cal O}}
 
 \def\cQ{{\cal Q}}
 \def\cR{{\cal R}}
 \def\cS{{\cal S}}

 \def\cV{{\cal V}}

 \newfont{\goth}{eufm10 scaled \magstep1}

 \def\gg{\mbox{\goth g}}

 \def\gl{\mbox{\goth l}}

 \def\gp{\mbox{\goth p}}

 \def\gs{\mbox{\goth s}}

 \def\ggl{\gg\gl}\def\gsl{\gs\gl}\def\psl{\gp\gs\gl}\def\pgl{\gp\gg\gl}
 
 \def\a{\alpha}
 \def\b{\beta}
 \def\C{\Gamma}
 \def\d{\delta}\def\D{\Delta}
 \def\e{\epsilon}
 \def\f{\phi}\def\vf{\varphi}
 \def\h{\eta}

 \def\l{\lambda}
 \def\m{\mu}
 
 \def\p{\pi}
 
 \def\r{\rho}
 
 \def\t{\tau}
 \def\th{\theta}

 \def\adt{\alpha'}
 \def\bdt{\beta'}

 \def\be{\begin{equation}}\def\ee{\end{equation}}
 \def\bea{\begin{eqnarray}}\def\eea{\end{eqnarray}}
 \def\ba{\begin{array}}\def\ea{\end{array}}

 \def\del{\partial}

 \def\unA{\underline A}
 \def\unB{\underline B}

 \def\str{\rm str}
 \def\xz{\times}

 \def\nab{\nabla}

\def\tcR{\tilde{\cR}}\def\tQ{\tilde{Q}}
\def\tm{\tilde{m}}

 \def\del{\partial}
 
 \def\bt{\bullet}
 \def\3dt{\dot{3}}
 
\def\half{{1\over2}}\def\qu{{1\over4}}

 \let\la=\label

 \let\bm=\bibitem{}

 \def\nn{\nonumber}
 \def\bd{\begin{document}}
 \def\ed{\end{document}}
 \def\bea{\begin{eqnarray}}
 \def\ba{\begin{array}}\def\ea{\end{array}}
 \def\eea{\end{eqnarray}}
 \def\ft#1#2{{\textstyle{{\scriptstyle #1}\over {\scriptstyle #2}}}}
 \def\fft#1#2{{#1 \over #2}}
 \newcommand{\eq}[1]{(\ref{#1})}
 \def\eqs#1#2{(\ref{#1}-\ref{#2})}
 \def\det{{\rm det\,}}
 \def\tr{{\rm tr}}\def\Tr{{\rm Tr}}
  \def\str{{\rm str}} \def\diag{{\rm diag}}
 \def\sdet{{\rm sdet}}
 \newcommand{\hoch}[1]{$^{#1}$}
\begin{document}

 \thispagestyle{empty}

 \hfill{KCL-TH-03-12}

  \hfill{\today}

 \vspace{20pt}

 \begin{center}
 {\Large{\bf Aspects of $N=4$ SYM}}
 \vspace{30pt}

 {P.J. Heslop\hoch{1,2} and P.S. Howe\hoch3}\\[20pt]

 {\sl \hoch1 II. Institut f\"ur Theoretische Physik
der Universit\"at Hamburg}\\

{\sl \hoch2 Institut f\"ur Theoretische Physik
der Universit\"at Leipzig}\\
 
{\sl \hoch3 Department of Mathematics,
 King's College, London}

 \vspace{60pt}

 \end{center}

 {\bf Abstract}

The properties of gauge-invariant composite operators and their
correlation functions in $N=4$ SYM are discussed in the analytic
superspace formalism.  A complete classification of the different
types of operators in the theory is given.  Operators can be
either protected or unprotected according to whether they do not
or do have anomalous dimensions, and the analytic superspace
formalism allows one to identify which type a given operator is in
a straightforward manner. A simple discussion is given of the
behaviour of reducible multiplets at threshold. It is pointed out
that there is a class of ``semi-protected'' operators which do not
have anomalous dimensions but which do not necessarily have
non-renormalised three-point functions when the other two
operators in the correlator are protected, although two-point
functions of such operators are non-renormalised. A complete
discussion of superconformal invariants in analytic superspace is
given. The paper includes a modified discussion of the
transformation rules of analytic superfields which  clarifies the
$U(1)_Y$ properties of operators and correlation functions and, in
particular, explicit examples are given of three-point correlation
functions which violate this symmetry. A tensor, $\cE$, invariant
under $SL(n|m)$ but not under $GL(n|m)$, is introduced and used in
the discussion of $U(1)_Y$ and in the construction of invariants.

 {\vfill\leftline{}\vfill \vskip  10pt

 \baselineskip=15pt \pagebreak \setcounter{page}{1}

\section{Introduction}

Over the past few years there has been a substantial resurgence of
interest in four-dimensional conformal field theories,
particularly supersymmetric ones, largely inspired by the
Maldacena conjecture relating IIB string theory on $AdS_5\xz S^5$
to $N=4$ super Yang-Mills theory on Minkowski space, the conformal
boundary of $AdS_5$ \cite{maldacena}. The subject now has an
extensive literature and there are several review articles to
which we refer the reader for lists of
references~\cite{agmo,Klebanov:2000me,D'Hoker:2002aw}.

In this paper we discuss the properties of gauge-invariant
composite operators and their correlation functions in $N=4$
superconformal field theory, that is to say $N=4$ SYM, in four
dimensions in the setting of analytic superspace. There have been
many studies of such operators from various points of view but we
believe that the analytic superspace formalism has some
advantages. One is that holomorphic fields which transform under
irreducible representations of the isotropy subgroup of the
superconformal group which defines analytic superspace
automatically transform irreducibly under the full superconformal
group \cite{Heslop:2001zm}, and that all operators in the theory
can be expressed as such holomorphic fields. A second is that it
is easy to tell which operators will be protected in the quantum
theory simply by looking at the representations they transform
under and whether they can be written in terms of single trace 1/2
BPS operators (chiral primaries or CPOs) on analytic superspace
\cite{Heslop:2001dr}. We recall that, as well as short (BPS)
operators, there are protected operators whose lack of anomalous
dimensions was first deduced indirectly from the study of
correlations functions \cite{Arutyunov:2000ku}. In
\cite{Heslop:2001dr} it was noted that this phenomenon has a much
simpler explanation in terms of representation theory in that the
operators concerned are also subject to shortening conditions. The
analytic superspace formalism makes it clear precisely which
operators will remain short in the quantum theory. In the
classical theory there can also be operators which transform under
short representations but which turn out to be descendants of long
operators and which therefore acquire the anomalous dimensions of
the associated long operators in the quantum theory. This has been
studied in detail for the case of 1/4 BPS operators
\cite{Ryzhov:2001bp,D'Hoker:2003vf} and again can be understood
simply in the analytic superspace formalism. This aspect of the
theory can be looked at from the opposite point of view. One can
consider the limit in which the anomalous dimension of a given
long operator disappears. For some operators this limit results in
a reducible representation. This reducibility at threshold was
studied in great detail in \cite{Dolan:2002zh} but it can also be
understood very simply in analytic superspace.

The main technical development in this paper is an improved, and
hopefully clearer, discussion of the transformation of operators
in analytic superspace which places greater emphasis on the r\^ole
of the so-called $U(1)_Y$ ``bonus symmetry"
group~\cite{Intriligator:1998ig}. There is some ambiguity
concerning the representations of the isotropy group associated
with  long operators which is related to the violation of this
symmetry for certain two- and three-point correlators involving
such operators. We also draw attention to the existence of a class
of ``semi-protected" operators which are ambiguous in this sense
and whose three-point correlators with other protected operators
may not be non-renormalised. These operators, which can be either
series A or series B, have the property that they saturate one
unitarity bound but not both. Such representations were called
  ``intermediate short'' in~\cite{Ferrara:2000eb}.
A key r\^ole in the analysis of these operators and their
correlators
 is played by a tensor, which we call $\cE$, which is
invariant under $\gs\gl(2|2)$ but not under
$\gg\gl(2|2)$.\footnote{Invariant tensors of this type exist for
any $\gs\gl(n|m)$ algebra; they  can be thought of as analogues of
the $\e$-tensor in $\gs\gl(n)$.} In addition, one can use this
tensor to give a simple discussion of the $U(1)_Y$ properties of
superconformal invariants on analytic superspace.

The main properties of $N=4$ SYM that can be established using the
analytic superspace formalism are: 1) the protection of operators,
i.e. the lack of anomalous dimensions of operators transforming
under short representations, 2) non-renormalisation theorems for
two- and three-point functions of protected operators, 3)
free-form and non-renormalisation of extremal and next-to-extremal
correlators of protected operators and 4) the partial
non-renormalisation of four-point functions of 1/2 BPS
operators.\footnote{It should be noted that these results are
predicated on the assumption that the superconformal Ward
Identities hold in the quantum theory with composite operators; to
our knowledge there is no rigorous proof of this currently in the
literature. } Of course, point 1 has been discussed from other
points of view, see, for example, \cite{Andrianopoli:1998ut}, but
the analytic superspace formalism makes it clear precisely which
operators are protected and allows a complete classification of
all protected operators \cite{Heslop:2001dr}. For point 2 we note
that, although early derivations of results of this type were
given either from the AdS point of view \cite{fmmr,ads} or in
(mainly) perturbative field theory \cite{nonren,hsw}, the analytic
superspace formalism allows a compelling non-perturbative argument
for the non-renormalisation of two- and three-point functions of
protected operators on the field theory side \cite{ehw}. This
argument makes use of the reduction formula, introduced in
\cite{Intriligator:1998ig}, and can also be interpreted in terms
of $U(1)_Y$ symmetry as conjectured (and proved in the two-point
case) in the same paper. It is applicable to general protected
operators and not just the one-half BPS ones \cite{Heslop:2001gp}.
As we mentioned above, one has to take extra care when these
correlators involve semi-protected operators. Points 3 and 4 will
not be discussed in this paper (see \cite{D'Hoker:2002aw} for
references), although we again note that the analytic superspace
formalism can be applied to extremal correlators involving general
protected operators, not just one-half BPS superfields
\cite{Heslop:2001gp}.

The $N=4$ superspace approach to $N=4$ SCFT is on-shell in the
sense that the underlying $N=4$ SYM multiplet satisfies the field
equations. In fact, one has to use these equations to establish
the analyticity of operators such as the supercurrent. This
circumstance has led some authors to criticise this approach and
it is indeed the case that one cannot write down a path integral
nor carry out perturbative calculations in $N=4$ superspace.
However, the viewpoint we adopt is somewhat different. We are
interested in analysing the constraints on the full
non-perturbative theory due to superconformal invariance and we
can view the $N=4$ formalism as a way of packaging the outcome of
quantum calculations which can be carried out in any convenient
manner, for example, using components. In the component formalism
the superconformal algebra only closes on the fields modulo the
equations of motion and gauge transformations and the standard
gauge-fixing terms explicitly break supersymmetry. All of these
technical difficulties can be overcome using the BRST/BV formalism
for a combined BRST algebra which includes the BRST versions of
both gauge and superconformal transformations, the latter
involving space-time independent ghosts
\cite{Howe:1990pz,White:wu}. In the study of correlation functions
of gauge-invariant operators, however,  one expects to recover the
naive supersymmetry Ward identities. Moreover, the use of the
equations of motion should be valid provided that one avoids
coincident points. In principle, one might encounter difficulties
due to contact terms when one uses the reduction formula which
relates the derivative of an $n$-point function with respect to
the coupling constant to an $(n+1)$-point function which has one
integrated insertion of the on-shell action, but there is no
evidence to date that any such difficulties arise. Moreover, it
seems that the formalism correctly encodes the expected anomalies
in the supercurrent due to the fact these are local so that the
formally analytic expressions for the correlators have to be
regularised in order to handle quantities that are ill-defined at
coincident points.\footnote{For example, the three-point function
of three supercurrent operators \cite{hsw} is formally analytic
but still encodes, amongst other anomalies, the usual triangle
anomaly for the $SU(4)$ currents due to the regularisation that is
required when one has coincident points \cite{fmmr}.}

The organisation of the paper is as follows: in the next section
we review the analytic superspace formalism and discuss the
transformations of operators. The isotropy algebra includes two
$\gs\gl(2|2)$ algebras which are analogous to the two $\gs\gl(2)$
spin algebras in ordinary Minkowski space. However, tensor fields
transform under two $\gg\gl(2|2)$ algebras from which one cannot
remove the supertrace in a canonical fashion. It is therefore
natural to use $\gg\gl(2|2)$ representations and Young tableaux to
describe tensor operators. These are not uniquely determined for
long operators so that our presentation of the theory differs
slightly from our earlier work which was largely concerned with
protected operators. In section three we classify the operators of
$N=4$ SYM with examples of each type of operator and discuss
reducibility at threshold, and comment briefly on mixing for
one-quarter BPS and other operators. In section four we discuss
the $U(1)_Y$ behaviour of two-and three-point functions. In
particular, we exhibit explicit examples of correlators which are
not invariant under this symmetry and which should therefore be
subject to renormalisation effects.  We show that the two-point
functions of a semi-protected operator and its conjugate are
non-renormalised even though there is a violation of $U(1)_Y$
symmetry in the three-point function involving an additional
supercurrent.  We also compare our results with the earlier
conjectures on $U(1)_Y$ made in reference
\cite{Intriligator:1999ff}.  In section five we give a complete
discussion of invariants in analytic superspace, both in
coordinate language and in the Grassmannian formalism of
\cite{Howe:1996rk}. The $\cE$ tensor is used to give a simple
explanation of the existence of superconformal invariants which
are not $U(1)_Y$ invariant. We give our conclusions in section
eight. In an appendix we show how to explicitly convert a
superconformal field on analytic superspace into a field on
harmonic superspace.

\section{Analytic superspace}

Analytic superspaces were introduced in \cite{Galperin:1984av};
for a review see \cite{Galperin:uw}.  A harmonic superspace is a
product of ordinary Minkowski space and a compact complex coset
space of the internal symmetry group. A field on harmonic
superspace which is both Grasmmann analytic (generalised chiral)
and analytic with respect to the internal manifold  can be written
as an unconstrained superfield on analytic superspace which has a
reduced number of odd coordinates. This is similar to the way a
chiral superfield can be written as an unconstrained superfield on
chiral superspace. The general theory of such superspaces realised
as coset spaces of complexified superconformal groups was
developed in \cite{Howe:md,Hartwell:1994rp} (see also
\cite{Lukierski:1988vw}) and applied to
four-dimensional super Yang-Mills theories in an earlier series of
papers summarised in \cite{Howe:2001je}.

The analytic superspace we shall be using in this paper is a coset
of the complexified $N=4$ superconformal group $PSL(4|4)$ with a
parabolic isotropy group. We recall that a parabolic subgroup is
one which contains the Borel subgroup. For the $SL(n)$ Lie groups
the latter can be thought of as the group of lower triangular
$n\xz n$ matrices, but in the supersymmetric case the Borel
subgroup is no longer unique (up to conjugation). The group
$(P)SL(4|N)$ acts naturally from the left on $\bbC^{4|N}$, and the
Borel subgroup can again be identified with the lower triangular
matrices, but one obtains inequivalent Borel supergroups for
different orderings of the basis elements of $\bbC^{4|N}$ with
respect to Grassmann parity. The most convenient choice for
applications to superconformal field theory has the form $(2|N|2)$
(i.e. $(2\ {\rm even}|N\ {\rm odd}|2\ {\rm even})$). For this
choice, the parabolic subgroup which defines the analytic
superspace we are interested in ($(4,2,2)$ analytic superspace)
consists of supermatrices of the form:

\be
 \left( \ba{cc|cccc|cc}
\bt &\bt &\bt & \bt&&&&\\
\bt &\bt &\bt &\bt&&&&\\
 \hline
\bt &\bt &\bt &\bt&&&&\\
\bt &\bt &\bt &\bt&&&&\\
\bt &\bt &\bt &\bt&\bt&\bt&\bt&\bt\\
\bt &\bt &\bt &\bt&\bt&\bt&\bt&\bt\\
\hline
\bt &\bt &\bt &\bt&\bt&\bt&\bt&\bt\\
\bt &\bt &\bt &\bt&\bt&\bt&\bt&\bt\\
\ea \right) \la{isot}
 \ee

where the bullets denote matrix elements which can be non-zero.
The blank spaces can be thought of as representing local
coordinates for analytic superspace, and these are therefore
described  by four sets of $2\xz 2$ matrices, two of which have
even elements and two of which have odd elements. As can be seen
from the above diagram, the coordinate supermatrix is not actually
in standard form so that it is convenient to make another
transposition of the basis (corresponding to the ordering
$(2|2|2|2)$ of $\bbC^{4|4}$) in order to effect this. We then
denote the coordinates by

 \be
 X^{AA'}=\left( \ba{cc} x^{\a \a'} & \l^{\a a'} \\
                        \pi^{a\a'}  &  y^{a a'}  \ea
 \right).
 \ee

However, we shall adhere to the $(2|4|2)$ notation when labelling
representations as this is more convenient for Minkowski
superspace and also because this is the notation we have used in
previous papers.

Geometrically, $(4,2,2)$ analytic superspace is the Grassmannian
of planes of dimension $(2|2)$ in $\bbC^{4|4}$. Fields on this
space are naturally what one might call generalised spinor fields,
i.e they carry $A$ and $A'$ indices which are acted on in a linear
way by the Levi subgroup of the isotropy group. This group
consists of block diagonal elements of the type given in \eq{isot}
(where the blocks have dimension $(2|2)\xz(2|2)$). This is
analogous to fields on Minkowski space carrying primed and
unprimed spinor indices . In order to obtain representations of
$PSL(4|4)$ it turns out that the fields must carry the same number
of primed and unprimed indices, while the representations will be
unitary if all the indices are downstairs (covariant). The Levi
subalegbra is $\gp\gs(\gg\gl(2|2)\oplus\gg\gl(2|2))$, but tensor
fields will transform naturally under the two $\gg\gl(2|2)$s
(acting on primed and unprimed indices). Since the unit $(n|n)\xz
(n|n)$ matrix has vanishing supertrace, it is not possible to
remove the supertrace from these algebras in an invariant way, and
so it is convenient to regard the tensor indices as $\gg\gl$
indices, despite the fact that the fields are labelled by
representations of the two $\gs\gl$ subalgebras and an additional
charge (the overall $P$ being taken care of by having equal
numbers of primed and unprimed indices as we have remarked above).
In general there will be many $\gg\gl(2|2)$ representations which
correspond to the same $\gs\gl(2|2)$ representation and this will
play a r\^ole in the discussion of the $U(1)_Y$ properties of
various correlators to be discussed below. Before studying fields
and correlators on analytic superspace we make a few remarks about
ordinary four-dimensional (complex) Minkowski space which is again
a Grassmannian; it is the space of  2-planes in $\bbC^4$.

\subsection{Minkowski space}

Complexified Minkowski space can be represented as a coset space
of $SL(4)$, where $SL(4)$ is the complexified conformal group,
with isotropy group consisting of matrices of the form

\be
 \left( \ba{cccc}
\bt &\bt &&\\
\bt &\bt &&\\
\bt &\bt &\bt &\bt\\
\bt &\bt &\bt &\bt\\
\ea \right). \ee

This space has coordinates $x^{\a \adt}$. The block-diagonal
(Levi) subalgebra of the corresponding Lie algebra is
$\gs(\gg\gl(2)\oplus\gg\gl(2))\sim
\gs\gl(2)\oplus\gs\gl(2)\oplus\bbC$.\footnote{In real Minkowski
space the two $\gsl(2)$ groups become complex conjugates of each
other and $\com$ becomes $\real$.} Conformal fields on Minkowski
space have various $\gs\gl(2)$ indices $\a, \adt$ and a dilation
weight $L$ which specify their transformation properties under the
isotropy group. For irreducible representations of the isotropy
algebra, only the block diagonal part, the Levi subalgebra, acts
non-trivially.  If we denote an element of $\d g\in\gs\gl(4)$ by

 \be
 \d g=\left(
 \ba{rr}
 -A^\a{}_\b & B^{\a \bdt}\\
 -C_{\adt \b} & D_{\adt}{}^{\bdt}
 \ea\right)
 \ee

then a conformal field $\cO(x)$ will transform in the following
way:

\be \d \cO= \cV\cO + \cR(A(x))\cO + \cR'(D(x))\cO + Q\D\cO \ee

where

\bea ({\cV} x)^{\a\adt}&=& B^{\a\adt} + A^{\a}{}_{\b} x^{\b\adt} +
x^{\a\bdt} D_{\bdt}{}^{\adt} +
x^{\a\bdt} C_{\bdt\b} x^{\b\adt}\nn\\
A(x)^{\a}{}_{\b}&=&A^{\a}{}_{\b} + x^{\a\bdt} C_{\bdt\b}\nn\\
D(x)_{\adt}{}^{\bdt}&=&D_{\adt}{}^{\bdt} + C_{\adt\a} x^{\a\bdt}\nn\\
\D&=& \tr(A +xC)=\tr(D+Cx)\,. \eea

$\cR$ and $\cR'$ denote representations of the two $\gg\gl(2)$
algebras, and $Q=L-(J_1+J_2)$ where $J_1$ and $J_2$ are the spin
quantum numbers of the two $\gsl(2)$ algebras. It is assumed that
the representations $\cR$ and $\cR'$ are given by Young tableaux
with only one row, i.e they correspond to $\gs\gl(2)$
representations. Such a field (provided that it satisfies
appropriate differential constraints) will correspond to an
irreducible highest weight representation of $\gs\gl(4)$. A useful
way of specifying such representations of simple (super) Lie
groups is by assigning a number to each node of the corresponding
(super) Dynkin diagram. These numbers give the highest weight of
the representation which in turn uniquely specifies the entire
irreducible representation. Furthermore one can specify a
parabolic coset space by putting crosses through some nodes of the
Dynkin diagram \cite{Baston}.\footnote{For $\gs\gl(n)$ the crosses
reduce the algebra to a direct sum of simple subalgebras and a
number of abelian subalgebras, one for each cross; this diagram
then represents the Levi subalgebra directly and the parabolic is
obtained by filling out block diagonal elements of $\gs\gl(n)$ to
block lower-triangular matrices} The resulting diagram can then be
used to read off the transformation properties of the field on
this space which carries the representation of the (super)
conformal group in question.

In complex Minkowski space, therefore, representations of the
complex conformal group $\gs\gl(4)$ can be encoded in the
following diagram:

\be
 \begin{picture}(30,10)(0,0) \put(0,0){\makebox[0pt][l]{$\bt
\hspace {2em}
 \xz \hspace{2em}\bt$} \rule[.5ex]{5.5em}{.1ex} }
 \put(-5,10){ $n_1$ \hspace{1em} $n_2$ \hspace {1em} $n_3$}
 \end{picture}
\ee

Here the three nodes are the Dynkin diagram for $\gs\gl(4)$, the
numbers above the nodes specify the representation we are
interested in, and the cross through the central node tells us we
are interested in Minkowski space. If we wish to find out which
field on Minkowski space carries this representation we can read
this directly from the diagram. As we have already noted, to
specify a conformal field on Minkowski space one needs to specify
the representation of $\gs \gl(2) \oplus \gs \gl(2) \oplus \com$
which acts linearly on the field. But the crossed through node
splits the Dynkin diagram up into two $\gs\gl(2)$ Dynkin diagrams
(consisting of single nodes) and a crossed through node. The
numbers above each single node gives the representation of each
$\gs\gl(2)$ and the number above the crossed node gives the $\com$
charge. Specifically, the above diagram corresponds to a field
with $n_1=2J_1$ symmetrised unprimed spinor indices, $n_3=2J_2$
symmetrised primed spinor indices, and dilation weight $L=-n_2
-1/2(n_1 +n_3)$. In addition, fields on Minkowski space must
satisfy differential constraints in order to carry irreducible
representations of the conformal group. For example the
representation with Dynkin labels $n_1=n_3=0,\ n_2=-1$ corresponds
to a massless scalar field and therefore satisfies the massless
Klein-Gordon equation.

A given $\gs\gl(2)$ tableau with $m$ boxes can be described in
$\gg\gl(2)$ by any two-row tableau of the form $<m_1 m_2>$ where
$m_1$ and $m_2$ specify the number of boxes in the second and
first rows of the $\gg \gl(2)$ Young tableau respectively (this
unusual notation ties in with the supersymmetric case later), and
where $m_2-m_1=m$. This means that, instead of using the
representations $\cR$ and $\cR'$ corresponding to single-row
tableaux with $n_1$ and $n_3$ boxes respectively, we could
equivalently use a field specified by the labels $(<m_1 m_2><m'_1
m'_2> \tilde Q)$, where  we use two-row Young tableaux to specify
the representations of the two $\gg\gl(2)$ algebras. In order to
describe the same representation as before we must  have
$m_1-m_2=n_1$, $m'_1-m'_2=n_3$ while $\tilde Q=Q +(m_1 +m'_1)$. In
other words, we may use a field which transforms under any
representations $\tilde{\cR},\tilde{\cR'}$ which correspond to the
same representations as $\cR,\cR'$ under the $\gs\gl(2)$
subalgebras of the $\gg \gl(2)$s, provide that we change the value
of the charge $Q$ appropriately.

This is a rather trivial manoeuvre in the present example, but it
will be useful in the $N=4$ case as we shall see below. The main
reason for the difference is that $N=4$ analytic superspace also
carries an action of $U(1)_Y$, and the charge of a given field
under this group will be different for different choices of
$\ggl(2|2)$ tensor representations.

\subsection{Fields on analytic superspace}

Representations of $SL(4|N)$ are specified by the quantum numbers
$(L,R,J_1,J_2,a_1,\dots,a_N)$ where $L$ is the dilation weight,
$R$ is the R-charge, $J_1$ and $J_2$ are spin labels and
$(a_1,\dots,a_N)$ are $SL(N)$ Dynkin labels. In $N=4$,
representations of $PSL(4|4)$ are representations of $SL(4|4)$
which satisfy the constraint $R=0$. They are therefore specified
by the quantum numbers $(L,J_1,J_2,a_1,a_2,a_3)$. The Dynkin
diagram for superfields on analytic superspace carrying such
representations of the $N=4$ superconformal group is

 \be
 \begin{picture}(192,10)(0,0) \put(0,0){\makebox[0pt][l]{$\bt
 \hspace {2em}
 \ominus \hspace{2em}\bt\hspace{2em}\times\hspace{2em}\bt
 \hspace{2em}\ominus \hspace{2em}\bt$}
 \rule[.5ex]{17.1em}{.1ex} }
 \put(-5,10){ $n_1$ \hspace{1em} $n_2$ \hspace {1em} $n_3$
 \hspace {1.5em} $n_4$\hspace {1.5em} $n_5$\hspace {1em} $n_6$
 \hspace {1em} $n_7$}
 \end{picture}\la{an}
 \ee

We see that the single cross splits the Dynkin diagram into two
smaller Dynkin diagrams which  each represent $\gs \gl(2|2)$, the
crossed-through node again representing a $\com$ charge. The white
nodes correspond to odd roots and the number and position of such
nodes depends on the choice of basis of $\bbC^{4|4}$ that has been
chosen; the one we are using here corresponds to the ordering
$(2|N|2)$. The two $\gsl(2|2)$ sub-Dynkin diagrams have a single
central white node and correspond to the usual basis ordering
$(2|2)$ of $\bbC^{(2|2)}$.

The quantum numbers are related to the super Dynkin labels by

\be
\ba{rcl} n_1 &=& 2J_1\\
             n_2 &=& {1\over2}(L-R) + J_1 + {M\over 4} -M_1\\
             n_{2+i}&=&a_i \qquad (i=1 \dots 3)\\
             n_{6}& =& {1\over2}(L+R)+J_2-{M\over 4}\\
             n_{7}&=&2J_2
               \ea \la{dynk} \ee

where $M$ is the total number of boxes in the Young tableau of the
internal $\gs\gl(4)$ representation, and $M_1$ is the number of
boxes in the first row of this tableau, i.e.

 \be
M=\sum_{k=1}^{3} k\,a_k\qquad  M_1=\sum_{k=1}^{3} a_i \ee

From \eq{dynk} we see that

\be R=\half(n_1-2n_2-n_3)+\half(n_5+2n_6-n_7) \ee

so that the representations that we are interested in, which have
$R=0$, satisfy the constraint

\be I:=n_3 + 2n_2 -n_1=n_5+2n_6-n_7. \la{r0}\ee

We shall see in a moment that this corresponds to tensors on
analytic superspace which have the same total number, $I$, of
primed and unprimed indices.  In terms of the Dynkin labels the
dilation weight $L$ is given by

\be L=\sum_{i=2}^6 n_i - (n_1+n_7) \ee

In the field theory context we are interested in unitary
representations (of the real superconformal group) \cite{screp};
there are three series of such operators (for general $N$) and
they satisfy certain unitarity bounds:

\be
 \ba{lll}
 {\rm Series\ A:} & L\geq 2+2J_1 + 2M_1 -{m\over2}\qquad &L\geq
 2+2J_2 +{M\over2} \\
 &&\\
 {\rm Series\ B:} & L= {M\over2};\ & L\geq 1+M_1 + J_1,\  J_2=0\ \
 {\rm
 or}\\
 &&\\
 &L=2m_1-{M\over2}; & L\geq 1 + M_1 + J_2,\ J_1=0\\
 &&\\
 {\rm Series\ C:} & L=M_1={M\over2} &J_1=J_2=0
 \ea
 \la{1}
 \ee

For $N=4$, these bounds can be rewritten in terms of the Dynkin
labels as follows:

\be \ba{rrclrcl} {\rm Series\ A}: & n_2&\geq &n_1 +1,\qquad&
n_{6}&\geq&n_{7}+1
\hspace{4cm}\\
&&&&&\nn\\
{\rm Series\ B}:& n_2&\geq &n_1 +1,& n_{6}&=&0, \qquad n_{7}=0\\
&&&&&\nn\\
 {\rm Series\ C}:& n_2&=&0,&n_{6}&=&0, \qquad
n_1=n_{7}=0 \ea \label{bounds} \ee

(or $n_1 \rightarrow n_{7}$, $n_2 \rightarrow n_{6}$ for series
B.)

These  bounds will be satisfied if the tensor representations of
the two $\gg\gl(2|2)$ algebras carried by the superfields
correspond to proper Young tableaux (see below) for covariant
tensors.

As in the case of Minkowski space, if we are given a
representation of the superconformal group, we can find a
corresponding analytic superfield which carries that
representation: the Dynkin labels specifying the representation of
$\gs\gl(4|4)$ in question also specify the representation of $\gs
\gl (2|2) \oplus \gs \gl(2|2) \oplus \com$ which the superfields
carry linearly (in practice as tensors with superindices). To
carry this out we need to know how to convert from $\gs\gl(2|2)$
Dynkin labels to Young tableaux so that we can explicitly write
down the tensor fields. However, as we remarked above, it is
easier to consider the tensor indices $A,A'$ as $\gg\gl(2|2)$
indices, and so we shall use Young tableaux for this bigger group.

An important difference between fields on Minkowski space and
fields on analytic superspace is that on analytic superspace all
holomorphic, irreducible tensor fields automatically carry
irreducible representations; no differential constraints need to
be imposed as in the case of fields on Minkowski space
\cite{Heslop:2001zm}. This is very similar to what happens in
twistor theory where irreducible representations of the conformal
group (for example massless fields in Minkowski space) are given
by cohomology classes on twistor space without constraints
(see~\cite{Baston}). In the supersymmetric case, however, one only
needs the zeroth cohomology classes (i.e. holomorphic superfields)
on analytic superspaces. This property enables one to write down
solutions to the superconformal Ward identities without having to
solve differential equations.

\subsection{$\gg\gl(2|2)$ Young tableaux}\la{sec:YT}

Superconformal operators on $(4,2,2)$ analytic superspace have
superindices which can carry either $\gsl(2|2)$ or $\ggl(2|2)$
representations (as compared to conformal operators on Minkowski
space which have spinor indices carrying representations of
$\gsl(2)$ or $\ggl(2)$). Representations of the left $\gsl(2|2)$
algebra are determined by the three Dynkin labels $(n_1,n_2,n_3)$
and those of the right $\gsl(2|2)$ algebra by the three labels
$(n_7,n_6,n_5)$ corresponding to the left and right halves
respectively of the diagram~\eq{an}.

As in the case of purely bosonic simple Lie algebras, all finite
dimensional representations of $\gsl(2|2)$ with integer
coefficients can be obtained by tensoring together copies of the
fundamental and/or anti-fundamental representation (which are
inequivalent for supergroups) and taking certain (anti-)symmetric
combinations which are given in a Young tableaux
\cite{Bars:1982se}. The Young tableaux is interpreted just as for
purely bosonic groups (the number of boxes corresponds to the
number of indices, rows correspond to symmetrised indices and
columns to anti-symmetrised) with the difference that the terms
symmetrised and anti-symmetrised are generalised to take into
account of the fact that  some of the indices are odd. This
implies that there is no limit to the number of boxes in a given
column. However, the tableau in the form of a $3\xz 3$ square is
identically zero, for symmetry reasons, and so  the tableaux are
restricted to have the form given below, with only the two
left-most columns having arbitrary length. We note that the tensor
product of representations can be computed by multiplying tableaux
together. The rules for this are the same as in the bosonic case,
but any tableau in the product which contains a $3\xz 3$ square
sub-tableau can be discarded.

The unitarity conditions \eq{1} imply that the unitary
representations are those obtained by tensoring the
anti-fundamental representation of $\gsl(2|2)$. However, as we
remarked above, it is convenient to work with $\ggl(2|2)$
tableaux. The most general such Young tableau that can be obtained
has the form:

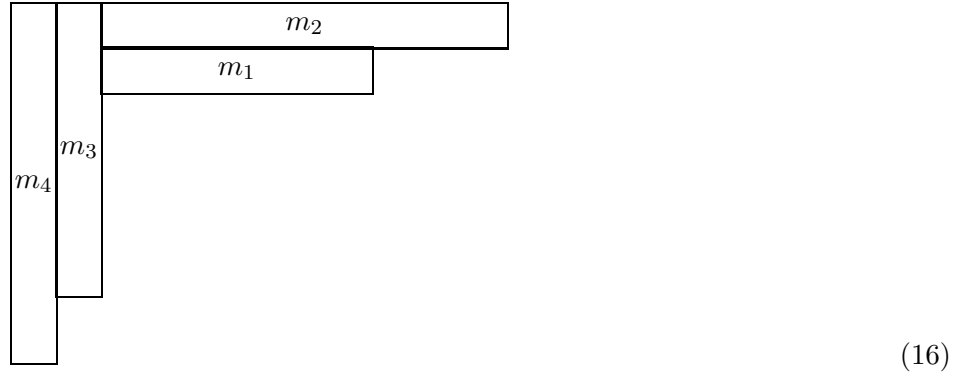
\be \setlength{\unitlength}{.3mm}
\begin{picture}(420,160)(0,0)
\put(100,140){\framebox(180,20){$m_2$}}
\put(100,120){\framebox(120,20){$m_1$}}
\put(60,0){\framebox(20,160){$m_4$}}
 \put(80,30){\framebox(20,130){$m_3$}}
\end{picture}
 \ee

where $m_i$ denotes the number of boxes in the indicated sections
of the Young tableaux. We will denote this Young tableau by
$<m_1,m_2,m_3,m_4>$. A diagram will be said to be proper if
$m_1\leq m_2,\ m_3\leq m_4$ and if $m_2\neq 0,\ m_3\geq 1$, while
if $m_1,m_2\neq 0,\ m_3\geq 2$. This diagram is related to the
Dynkin labels $\{n_i\}$ by

\be \ba{rcl}
m_4-m_3   &=&n_3\\
m_3+m_2   &=&n_2\\
m_2-m_1   &=&n_1. \ea \la{DY}\ee

We get a similar Young tableau $<m'_1,m'_2,m'_3,m'_4>$ for the
right $\gsl(2|2)$ algebra with

\be \ba{rcl}
m'_4-m'_3   &=&n_5\\
m'_3+m'_2   &=&n_6\\
m'_2-m'_1   &=&n_7. \ea \ee

Since there are only three numbers which determine each
representation of $\gsl(2|2)$ and there are four numbers
determining the Young tableau, in general we can have different
tableaux giving the same $\gsl(2|2)$ representation. This must be
the case as the Young tableaux also give representations of
$\ggl(2|2)$ and different $\ggl(2|2)$ representations may
correspond to the same $\gsl(2|2)$ representation (just as in the
case of Minkowski space we can have different $\ggl(2)$
representations corresponding to the same $\gsl(2)$
representation). Specifically,

\be <m_1,m_2,m_3,m_4> \sim <m_1-m,m_2-m,m_3+m,m_4+m> \la{sim}\ee

for any integer $m$, such that $2-m_3 \leq m \leq m_1$ (here
`$\sim$' means `corresponds to the same $\gsl(2|2)$ representation
as'.) In particular for any $\gsl(2|2)$ representation there is
always a Young tableau with $m_1=0$ corresponding to this
representation. We refer to this choice of Young tableau as the
canonical form of the representation.

One way of interpreting this is in terms of a tensor which is
invariant under $\gsl(2|2)$ but not under $\ggl(2|2)$. This can be
considered to be an analogue of the $\e$-tensor. In $\gsl(2)$ the
tableau $<1,1>$ (two boxes in a column) is equivalent to the
trivial tableau, so that one can deduce the existence of a
two-index antisymmetric tensor which is invariant under $\gsl(2)$,
but not under $\ggl(2)$. In $\gsl(2|2)$ the tableau $<1,1,2,2>$
(two rows each with three boxes) is equivalent to the tableau
$<0,0,3,3>$ (two columns each with three boxes), so that there is
a tensor with 6 covariant indices in the symmetry pattern of the
first tableau and 6 contravariant indices in the symmetry pattern
of the second tableau which is invariant under $\gsl(2|2)$ but not
under $\ggl(2|2)$; it transforms by a factor of the
superdeterminant under the bigger group. We  shall refer to this
tensor as $\cE$. There is also an inverse tensor for which the
covariant and contravariant symmetry patterns are interchanged. If
we start from a tensor in a representation with $m_1=0$ we can
obtain a tensor in the representation with $m_1=m$ by applying
$\cE$ $m$ times (together with appropriate projections). We shall
write this operation schematically as $\cE(m)$.

As well as the tensor representations of $\gsl(2|2)$, with integer
Dynkin labels $[a,(a+b),c]$, there are also representations with
$b$ non-integral. We call these quasi-tensor representations. In
the superconformal field theory context unitarity implies $b$ must
be real and greater than one, with the limit $b=1$ (discussed in
subsection 3.2) giving rise to reducible representations. For
simplicity we can think about the case $[0,b,0]$. For $b$ a
positive integer the canonical Young tableau for this
representation is $<0,0,b,b>$, i.e. a two-column tableau with $b$
boxes in each column. The other possible tableaux will be of the
form $<k,k,b-k,b-k>$. The representation with $b=1$ is short but
all of the others, with $b\geq 2$, have the same dimension. We can
therefore realise such representations on fields which may be
chosen to have the same index pattern as $<0,0,2,2>$. If $p=b+2$,
we shall denote such a field by $\cO[p,0]$. Instead of using the
canonical tableau we could have used, for example, the tableau
$<1,1,b-1,b-1>$. Again all such tensors have the same number of
components and can therefore be represented on fields which have
the same index pattern as the tableau $<0,0,2,2>$. We denote such
a field by $\tilde\cO[p-1,1]$. The two representation spaces are
isomorphic and are related by an isomorphism which we can again
denote by $\cE$, so $\tilde\cO[p-1,1]=\cE \cO[p,0]$. All of this
continues to make sense for $b$ real, $b>1$, so that we can extend
the notion of the $\cE$ tensor to the case of quasi-tensors.

\subsection{Field transformations in analytic superspace}

For any unitary irreducible representation we are now in a
position to be able to give a superfield on analytic superspace
which carries this representation. If the representation has
Dynkin labels $[n_1n_2\dots n_7]$ (which can be obtained from the
usual quantum numbers from \eq{dynk}) then we denote by $\cR,\cR'$
the left and right representation of $\gsl(2|2)$ respectively. In
practice these will be specified by tensor indices, symmetrised
according to Young tableaux as described above. There is some
ambiguity as to how to do this, so we will choose the canonical
Young tableau, with $m_1=0$, that is

\bea \cR &=& <0,n_1,n_2-n_1,n_3+n_2-n_1> \la{can}\\
\cR' &=&<0,n_7,n_6-n_7,n_5+n_6-n_7>
 \eea

The total number of indices of the representation $\cR$, given by
the number of boxes in the corresponding Young tableau, is

\be I:=n_3 + 2n_2 -n_1=n_5+2n_6-n_7. \ee

This equality occurs because we are only considering
representations of $SL(4|4)$ with zero R-charge (equivalent to
studying representations of $PSL(4|4)$) and this translates into
the condition that the number of indices of the $\cR$ and $\cR'$
representations is the same.

We denote a general operator by $\cO_{\cR \cR'}^Q$ where
$Q=L-J_1-J_2$ for canonical tableaux. Under an infinitesimal
superconformal transformation specified by an infinitesimal $\gs
\gl(4|4)$ matrix

 \be
 \d g=\left(
 \ba{rr}
 -A^A{}_B & B^{A B'}\\
 -C_{A' B} & D_{A'}{}^{B'}
 \ea\right),
 \ee

where each entry is a $\gg \gl(2|2)$ matrix and where
$\str(A)=\str(D)$, an operator $\cO_{\cR \cR'}^Q$ transforms as

\be
 \d \cO_{\cR \cR'}^Q = \cV \cO_{\cR \cR'}^Q + \cR(A(X)) \cO_{\cR
 \cR'}^Q + \cR'(D(X)) \cO_{\cR \cR'}^Q + Q\D \cO_{\cR \cR'}^Q.
 \la{trans}
 \ee

where

\bea
 \cV X &=& B + AX + XD + XCX \la{VX}\\
 A(X)&=&A + XC \\
 D(X)&=& D + CX\\
 \D&=&\str(A + XC)=\str(D+CX).\la{delta}
 \eea

Here $\cV$ is the vector field generating the transformation.
$A(X)$ and $D(X)$ are $\gg \gl(2|2)$ matrices rather than $\gs
\gl(2|2)$ matrices, but the Young tableaux define representations
of this group as well and so $\cR(A(X)),\ \cR(D(X))$ make sense.
Note that the unit $(4|4)\xz (4|4)$ matrix does not act on
analytic superspace, and if we only consider superfields that have
equal numbers of left and right superindices, then the identity
matrix does not act on the superfield indices  either so we
automatically obtain representations of $\psl(4|4)$.

\subsection{$U(1)_Y$ and $PGL(4|4)$ transformations}

Although the unit matrix does not act on analytic superspace there
is nevertheless a non-trivial action of the algebra $\pgl(4|4)$
which extends $\psl(4|4)$ by an abelian algebra. We shall refer to
the corresponding abelian group as $U(1)_Y$ even though in the
complexified setting it is really a $\bbC^*:=\bbC\setminus\{0\}$
group. In the free theory, this group is a symmetry, and we can
extend the $\psl(4|4)$ transformations to $\pgl(4|4)$
transformations.  For the infinitesimal transformations we are
considering this simply amounts to removing the condition $\str(\d
g)=0$ (that is, we drop the constraint $\str(A)=\str(D)$). The
transformation of $X$ looks exactly as before~\eq{VX}, but now the
matrices $A$ and $D$ are unrestricted.

If we consider an $\ggl(4|4)$ transformation given by a diagonal
matrix with

 \be
 A\sim {1\over 2}\left(\ba{cc}
 a_o I_{(2|2)}& 0\\
 0& a_1 I_{(2|2)}\ea\right)
 \la{39}
 \ee

where $I_{(2|2)}$ denotes the unit tensor acting on $\bbC^{2|2}$,
and similarly for $D$, we find the transformations

 \bea
 \d x&=& s x \\
 \d\l&=& {1\over 2}(s + s' + \D')\l\\
 \d\p&=& {1\over 2}(s + s' -\D')\l\\
 \d y&=& s' y
 \la{40}
 \eea

where

 \bea
 s&=& {1\over2} (a_o + d_o)\\
 s'&=& {1\over2} (a_1 + d_1)\\
 \D'&=&{1\over2}(a_o-a_1-(d_o-d_1))=-{1\over2}\str(\d g)
 \la{41}
 \eea

The parameters $s$ and $s'$ correspond to dilations and internal
dilations respectively, and we can identify $\D'$ as the $U_Y(1)$
parameter which acts only on the odd variables. Note that

\be \D'=\half(\str(A+XC)-\str(D+CX))=\half(\str(A-D)) \ee

There is some ambiguity in the extension of the definition of $\D$
(equation \eq{delta}) to this case. We shall take it to be

\be \D={1\over 2}(\str(A+XC)+ \str(D+CX)) \ee

This definition is different to the one used in our previous
papers, but has the advantage that the free field-strength
superfield can be assigned zero $U(1)_Y$ charge, $Q'$. We recall
that this superfield, which has quantum numbers
$(L,J_1,J_2,a_1,a_2,a_3)=(1,0,0,0,1,0)$, or, equivalently, super
Dynkin labels $[0001000]$, is represented on analytic superspace
as a single-component superfield $W$ with $Q=1$.  In other words,
$W$ transforms as

\be \d W = \cV W + \D W \la{dW}\ee

under a $\pgl(4|4)$ transformation. Here $\cV$ is the vector field
which generates $\pgl(4|4)$ transformations on analytic
superspace: it has the same form as \eq{VX} but with the
constraint $\str A=\str D$ dropped.

Let us now consider the transformation of an arbitrary operator
$\cO^{QQ'}_{\cR\cR'}(X)$ on analytic superspace under the extended
algebra $\pgl(4|4)$. The field is specified by the ten labels
$(<m_1 m_2 m_3 m_4>,$ $\ <m'_1 m'_2 m'_3 m'_4>, Q,Q')$, but one
only needs seven to specify a representation of this group. Since
the number of primed and unprimed indices are equal we have $\sum
m_i = \sum m'_i$, so that only 9 of the labels are independent.
Furthermore, we can alter the representations $\cR$ and $\cR'$
without changing the way the field transforms under the two
$\gsl(2|2)$ algebras, and still have the same representation
provided that the charges are also adjusted. Thus two more of the
labels are redundant and the representation is therefore specified
by seven quantum numbers as expected. The transformation of the
field is given by

\be \d \cO^{QQ'}_{\cR\cR'}=\left(\cV + \cR(A(X)) + \cR'(D(X)) +
Q\D + Q'\D'\right) \cO^{QQ'}_{\cR\cR'} \la{pgl}\ee

To see what happens when we change the $\ggl(2|2)$
representations, suppose that $\cR$ and $\tcR$ are two
representations given by the Young tableaux $<m_1 m_2 m_3 m_4>$
and $<\tm_1 \tm_2 \tm_3 \tm_4>$ respectively where
$\tm_1=m_1+m,\tm_2=m_2+m,\tm_3=m_3-m,\tm_4=m_4-m$, so that the two
representations correspond to the same representation in
$\gsl(2|2)$. If $A\in \ggl(2|2)$ one has

\be \tcR(A)=\cE(m)\cR(A)\cE(m)^{-1} +m\,\str\, A \la{+m}\ee

and so we find (with a similar change of representation for the
primed algebra),

\be \tcR(A(X)) + \tcR'(D(X)) + \tQ\D + \tQ'\D'\sim \cR(A(X)) +
\cR'(D(X)) + Q\D + Q'\D' \ee

where

\be \tQ=Q-(m+m');\qquad \tQ'=Q' - (m-m') \ee

So we can change the $\ggl(2|2)$ representations that a field
transforms according to, while preserving the $\gsl(2|2)$
representations, provided that we adjust the charges accordingly.

In the free $N=4$ SYM theory, any unitary representation can be
written on analytic superspace in terms of free Maxwell
superfields such as $W$ and derivatives $\del_{A'A}$ and will have
the schematic form $\del^I W^Q$. We can form irreducible operators
by projecting the $I$ primed and unprimed indices onto irreducible
$\ggl(2|2)$ representations $\cS,\cS'$, thereby obtaining an
operator which we denote by $\cO^Q_{\cS\cS'}$. It will transform
under $\pgl(4|4)$ transformations according to the formula
\eq{pgl} with $\cR,\cR'$ replaced by $\cS,\cS'$ and $Q'=0$. A
simple example of such an operator is the supercurrent $T:=\tr
(W^2)$; this has no indices and hence corresponds to trivial
tableaux, and since it has two powers of $W$ it has $Q=2$.  The
Konishi superfield, on the other hand, is given in the free theory
by

\be \cO_{AB,A'B'}=\del_{(A(A'}W
\del_{B)B')}W-{1\over6}\del_{(A(A'}\del_{B)B')}(W^2) \la{kon} \ee

where both pairs of superindices are symmetrised. So there are
$Q=2$ $W$s and it has left and right Young tableaux $<0,0,1,1>$.

For any such operator we can change the tableaux $\cS,\cS'$
determined by the symmetry properties of the indices to the
corresponding canonical tableaux $\cR,\cR'$ for which
$m_1=m'_1=0$; we can thereby obtain an equivalent operator which
will describe the same representation of $\pgl(4|4)$ provided that
we change the charges $Q$ and $Q'$. So we have

\be \cO^Q_{\cS\cS'}\sim \cO^{\tQ Q'}_{\cR\cR'} \ee

where, if $\cS$ corresponds to the tableaux $<m_1..m_4>$, $\cR$
corresponds to the tableau $<0,m_2-m_1,m_3+m_1,m_4+m_1>$, and
similarly for $(\cS',\cR')$, and where

\be \tQ=Q +(m_1+m'_1) ;\qquad Q'=m_1-m_1' \ee

Thus any tensor operator of this type has well-defined properties
under the group $PGL(4|4)$. We shall say that the operator
$\cO^Q_{\cS\cS'}$ has $U(1)_Y$ charge $m_1-m'_1$. In fact, the
$U(1)_Y$ change of the highest weight state is $m_1-m'_1
+(J_1-J_2)$.

This can be generalised to the interacting quantum theory. The
quantum numbers of a given operator will not change, with the
possible exception of the dimension which may become anomalous. So
the Young tableaux and $Q$ charge can be assigned in a
straightforward way, and again we will obtain operators with
$Q'=m_1-m'_1$. However, not all operators in the free theory can
be straightforwardly generalised to the interacting classical
theory as fields on analytic superspace. This is because  the
field strength superfield $W$ transforms under the adjoint
representation of the gauge group and is covariantly analytic. It
is therefore no longer a holomorphic field on analytic superspace.
Moreover, there is no notion of a gauge covariant derivative
$\nab_{A'A}$ on this space. The operators that cannot be written
in terms of gauge-invariant products of $W$s and derivatives of
these in the interacting theory can be either long operators like
the Konishi operator or descendants of long operators. Note that,
for example, the interacting Konishi operator can be written as a
field, component by component, on analytic superspace even though
it cannot be written in terms of derivatives and $W$s.

The protected operators are those which can be written in terms of
derivatives and gauge-invariant products of $W$s and which are in
shortened representations (possibly series A). Such operators
saturate a unitarity bound and have either $n_2=n_1+1$ or both
$n_1=n_2=0$ and/or  similar  constraints for $n_7,n_6$. When
constraints of this type are satisfied for either $\ggl(2|2)$
tableau then it is easy to verify that this tableau must be in
canonical form, i.e. $m_1=0$. For more general operators this is
not the case, but we can nevertheless bring the tableaux to
canonical form as long as we change the charges $Q$ and $Q'$. If
the original, naturally defined, operator has $m_1\neq m'_1$, then
the form of the operator with canonical tableaux will have
non-zero $U(1)_Y$ charge given by $m_1-m'_1$. However, as we shall
see below, there is nothing in principle to stop long operators
which transform under the same representation of $PSL(4|4)$, but
which have different $U(1)_Y$ charges, mixing in the quantum
theory, so that $U(1)_Y$ charge will only be a good quantum number
for the protected operators

\section{Classification of operators}

In this section we  summarise the results that can be established
for various operators in $N=4$ SYM using the analytic superspace
formalism. It is possible to write down explicitly all operators in
the free theory on analytic superspace as unconstrained
superfields and this perhaps provides the simplest way of finding the
full spectrum of
gauge invariant operators in the theory. It would be  very interesting to
compare this with recent results concerning the spectrum of string
theory in $AdS_5 \times S^5$~\cite{Bianchi:2003wx}.

\subsection{Classification}

We begin with the free theory. In this case there are no anomalous
dimensions and one can explicitly construct examples of operators
which transform according to any irreducible representation of the
superconformal group from products of free field strength tensors
and analytic superspace derivatives \cite{Heslop:2000mr}. If we
now suppose we have $N_c^2-1$ of these which transform under the
adjoint representation of a rigid $SU(N_c)$ symmetry group and we 
demand that our operators be invariant under this group, then not
all representations of the superconformal group will be
permissible.

Now let us consider the classical interacting theory where the
$SU(N_c)$ group is taken to be a gauge group and the operators are
required to be gauge-invariant. Among the operators listed above
there will be those which involve derivatives acting directly on
$W$s. However, since there is no gauge-covariant derivative on
analytic superspace, such operators will not generalise to the
interacting case as operators on analytic superspace. These
operators can be of two types: operators which become long, such
as the Konishi operator, and descendants of long operators which
will not exist as separate operators in the interacting quantum
theory. The remaining operators can be constructed from products
of the single trace 1/2 BPS operators and ordinary analytic
superspace derivatives. These operators can be either long or
short.

Finally, let us consider the interacting quantum theory. The
operators which were allowed on analytic superspace classically
will split into two groups: those which are subject to a
shortening condition which will be analytic tensor fields
constructed from derivatives and products of single-trace 1/2 BPS
operators, and those which are not short. The latter will not
saturate any unitarity bound and this means that there are nearby
representations with the same number of components but with
anomalous dimensions. Such operators are therefore unprotected and
will develop anomalous dimensions. In the quantum theory they will
therefore become quasi-tensor superfields on analytic superspace.
We note that the protected operators also divide into two classes:
there are those which satisfy two unitarity bounds and which, as
we have seen, have unique $\gg\gl(2|2)$ Young tableaux and there
are also some operators which satisfy only one unitarity bound.
These operators will have one fixed tableau and one which is
ambiguous.

It should be noted that there is a subtlety in the precise
definitions of the  components of some protected operators which
can mix with operators in the same classical representation but
which are descendants. This was observed in~\cite{Ryzhov:2001bp}
for the case of one-quarter BPS operators and further details were
given in \cite{D'Hoker:2003vf}. One anticipates that  a similar
phenomenon should occur for some series B and short series A
operators. This complication does not affect the classification we
have given above, although one might say that a more precise
statement is that there is a one-to-one correspondence between
operators which can be written on analytic superspace in terms of
derivatives and single-trace 1/2 BPS operators and protected
operators in the full quantum theory.

A possible explanation of this behaviour in perturbation theory is
as follows: in the free theory the descendant one-quarter BPS
operators are short multiplets and so can (and do) mix with
operators in the same representations which are not descendants.
However, as soon as the coupling is switched on the descendants
cease to exist as independent multiplets and become part of
irreducible long multiplets. Superconformal symmetry would
therefore imply that they cannot mix with the true one-quarter BPS
operators. One would therefore expect the mixing to occur only at
zeroth order in the coupling; there should not be any quantum
corrections. Indeed the quarter BPS operators found
in~\cite{D'Hoker:2003vf} have been shown to remain unmodified at order $g^2$~\cite{Beisert:2003tq}.

There are also operators such as the Konishi operator which are
reducible in the classical interacting theory but which become
irreducible in the quantum theory after they acquire anomalous
dimensions. In the quantum theory they can therefore be
represented by quasi-tensor superfields on analytic superspace. It
is easiest to see what happens to these operators in the classical
limit by switching off the anomalous dimension at which point we
find that the representation becomes reducible.

We now list the different types of gauge-invariant operators in
$N=4$ SYM and give some examples.

\subsubsection*{CPOs}

The simplest protected operators are the single-trace one-half BPS
operators (CPOs) which have the form $A_Q:=\tr (W^Q)$. These
operators are in one-to-one correspondence with Kaluza-Klein
states of IIB supergravity on $AdS_5\xz S^5$. This fact was
pointed out in \cite{Andrianopoli:1998jh}, although this family of
operators had been considered as analytic superfields previously
\cite{Howe:1995aq}. Indeed, one can explicitly derive the relation
between the supergravity and field theory multiplets directly in
superspace \cite{Heslop:2000np}. The operator $T:=A_2$ is special;
it is the supercurrent and is extra-short. It has independent
components up to fourth order in the odd variables. It also
contains all of the conserved currents in the theory. The operator
$A_3$ is also extra-short; it has components up to sixth order in
the odd coordinates but contains no conserved currents. All other
operators in the sequence are full single component analytic
superfields with independent spacetime components up to eighth
order in the odd coordinates.

\subsubsection*{One-half BPS}

As well as the CPOs one can also have multi-trace one-half BPS
states by taking products of the CPOs. These operators all have
Dynkin labels of the form $[000Q000]$ and are therefore given as
single-component superfields on analytic superspace. The simplest
example is $T^2$ which has charge 4 and hence is in the same
representation as $A_4$.

\subsubsection*{One-quarter BPS}

The other class of series C operators are the one-quarter BPS
states which have Dynkin labels $[00pqp00]$. These are represented
by analytic tensor superfields which have $p$ derivatives with
both the primed and unprimed indices totally anti-symmetrised.
(The Young tableau $<0,0,0,p>$ has $p$ boxes in a single column.)
If $p=1$, these operators are covectors on analytic superspace,
and the simplest example of this class is the operator $[0013100]$
which can be written explicitly as

 \be
 \cO_{A A'}=\del_{[A[A'}T \del_{B]B']}A_3 + ...
 \ee

where the dots denote further terms required to ensure that the
operator is a primary field. Note that these operators, and all
other tensor operators, involve derivatives and so must be
multi-trace. In the classical theory there are also operators
which transform under one-quarter BPS representations which are
single-trace. However, these are descendants of long operators and
are not BPS in the quantum theory. The simplest example of this
behaviour occurs for the representation $[0020200]$. There is a
double-trace BPS operator in this representation

 \be
 \cO_{AB A'B'}=\del_{[A[A'} T \del_{B]B']}T + ...
\ee

There is also a single-trace operator: in the free theory (but
with an $SU(N_c)$ group) this can be written

\be \cO_{ABA'B'}=\tr( [\del_{[A[A'} W, W] [\del_{B]B']} W,W] ) \ee

but this expression does not generalise to the interacting theory
because of the absence of a covariant derivative on analytic
superspace. A systematic study of these operators is  given in
\cite{Ryzhov:2001bp,D'Hoker:2003vf}.

\subsubsection*{Series B and series A protected}

A series B operator which saturates the series B unitarity bound
has Dynkin labels $[00 n_3 n_4 n_5 n_6 n_7]$ where $n_6=n_7+1$ and
$n_3=n_5+ n_7 + 2$ (or $[n_1 n_2 n_3 n_4 n_5 00]$ with $n_2=n_1+1$
and $n_5=n_3+n_1 +2$). The true (protected) series B operators in
this class are at least triple trace, but there can be descendants
in the classical theory which are single- or double-trace. The
scalar series B operators which saturate the unitarity bound have
Dynkin labels of the form $[00(q+2)pq 10]$. They have dimension
$L=2q+p+3$ which must be at least six as they are triple-trace.
The simplest example has $p=0$ and $q=3$, and therefore has
$SU(4)$ labels $[503]$ and $L=9$. It can be written as

 \be
 \cO_{ABCDE,A'B'C'D'E'}=\del_{AA'}T\del_{BB'}\del_{CC'}A_3
 \del_{DD'}\del_{EE'}A_4+\ldots
 \ee

where the unprimed indices are antisymmetrised, and the primed
indices are put in the $<0014>$ Young tableau pattern.

A series A protected operator has $n_2=n_1+1$ and $n_6=n_7+1$. The
simplest example of this type of operator has Dynkin labels
$[0102010]$; it is explicitly given by

 \be
 \cO_{AB,A'B'}=\del_{(A A'}T\del_{B)B'}T+...
 \ee

This is symmetric on both pairs of indices. The internal Dynkin
labels are $[020]$ and, since $J_1=J_2=0$, this corresponds to a
scalar operator on Minkowksi superspace. It is the square of the
supercurrent in the real $20$-dimensional representation of
$SU(4)$. The fact that this operator, and some other series A
operators, are protected was inferred from correlator results
\cite{Arutyunov:2000ku,Arutyunov:2001mh,Arutyunov:2001qw,Bianchi:1999ge}.
The simpler explanation in terms of representations was given in
\cite{Heslop:2001dr} where many other examples are discussed. Note
that the representation itself does not determine a protected
operator of this type; there can be other realisations which have
the same quantum numbers in the free theory but which become
reducible in the classical interacting theory and long in the
quantum interacting theory. For example, the representation
$[0102010]$ can also be realised as a product of the Konishi
operator and the supercurrent in the free theory. In the classical
interacting theory it is reducible, and can no longer be written
as a tensor field on analytic superspace because this would
require gauge-covariant analytic superspace derivatives acting
directly on $W$. In the full quantum theory this operator
transforms under the irreducible representation
$[0(1+b)020(1+b)0]$ and has anomalous dimension equal to $2b$.

\subsubsection*{Semi-protected operators}

These are series A operators which saturate one, but not both,
unitarity bounds or series B operators which do not saturate the
unitarity bound.
The simplest series B example has labels $[0040020]$. The
right-hand tableau is a $2\xz 2$ square while the left-hand one is
four boxes in a column. It can be realised explicitly as

\be \cO_{ABCD,A'B'C'D'}=\del_{AA'}T \del_{B B'}T
\del_{CC'}\del_{DD'}T+ \ldots \ee

where the unprimed indices are in the tableau $<0004>$ (totally
antisymmetric) and the unprimed indices are put into the
$<0,0,2,2>$ tableau. It is interesting to observe that the
right-hand $\gg\gl(2|2)$ representation belongs to the series of
long representations $\cR_{abc}$ described in the next subsection.
So, as far as this subgroup is concerned, there are nearby
representations with non-integer values of $n_2$. However, since
the left-hand side is in a short representation of the second
$\gg\gl(2|2)$, and since the left and right sides are related
because $R=0$, it follows that $n_2$ must remain integral in the
quantum theory and so the operator will be protected  (in the
absence of an anomaly).

An example of a series A semi-protected operator is given by
$[0202210]$. It can be realised  as

\be \cO_{ABCD,A'B'C'D'}= \del_{AA'} T \del_{BB'} T\del_{C
C'}\del_{D D'}A_3 +\ldots \ee

with the indices projected onto the representations specified by
the tableaux $<0022>$, for the unprimed indices, and $<0013>$ for
the primed indices.

\subsubsection*{Long operators}

The long operators are series A operators for which neither bound
is saturated. The simplest example is the Konishi operator. On
Minkowski superspace it is an unconstrained scalar superfield with
dimension $L=2+2b$. The Dynkin labels are $[0(1+b)000(1+b)0]$. To
view it as a superfield on analytic superspace it is useful to
consider first the representation $[0200020]$. This corresponds to
a tensor with the primed and unprimed indices both in the
representation with tableau $<0,0,2,2>$, i.e. a $2\xz 2$ square.
In section~\ref{harm} we shall show that this gives a scalar
superfield on super Minkowski space. Indeed, all of the two-column
tableaux $<0,0,k,k>,k\geq 2$ have the same number of components
and also correspond to scalar superfields on Minkowski superspace.
Moreover, one can take $k$ to be non-integral as long as $k>1$
without changing the number of components.

There are also long operators which can be represented as tensors
on analytic superspace in the classical interacting theory but
which become quasi-tensors in the quantum theory. These must be
multi-trace.  The simplest example is $[0200020]$. On analytic
superspace this is

\be \cO_{ABCD,A'B'C'D'}= \del_{[AA'}\del_{B]B'}
T\del_{[CC'}\del_{D]D'}T+ \ldots \ee

where both sets of indices are projected onto to the $2\xz 2$
square tableau representation. In real Minkowski superspace this
operator is simply the square of the supercurrent in the singlet
representation of $SU(4)$.

\subsection{Reducible representations}

Any unitary irreducible representation of the $N=4$ superconformal
group can be represented by a Dynkin diagram with labels
$[n_1,\ldots n_7]$, subject to the constraint \eq{r0}, as we have
seen. Moreover, one can find how to  describe the tensor fields
which carry these representations on any coset of the
superconformal group with parabolic isotropy group by crossing
through some of the nodes. As we have indicated, this procedure
specifies the parabolic subgroup and also determines the tensor
structure of the field which carries the given representation. In
general, such a field may be subject to further differential
constraints, as in the case of a scalar field on Minkowski space.
However, for analytic superspace this is not the case; the
representations are automatically irreducible by holomorphicity. A
general analytic superspace is characterised by the property that
it only has crosses through internal nodes, whereas harmonic
superspaces, Minkowski superspace and super twistor spaces have
crosses through the white nodes and/or the external black nodes
which correspond to the spacetime spin labels (for the basis
choice we have been using). This means that fields on these
superspaces may need to satisfy further constraints in order to
carry irreducible representations.

For some series A operators it can happen that a superfield in the
interacting classical theory describes a reducible, but not
completely reducible, representation. This situation corresponds
to the existence of descendant operators which transform under
short representations in the classical theory but which do not
exist as independent representations in the quantum theory due to
the presence of anomalous dimensions. That is to say, in the
quantum theory, the original long representation becomes
irreducible. This problem has recently been studied at great
length in \cite{Dolan:2002zh}; here, we show that it has a very
simple description in the analytic superspace formalism. It is
related to the notion of quasi-tensors discussed in
\cite{Heslop:2001gp}.

To illustrate the phenomenon, consider first the $N=2$ theory on
analytic superspace. The super Dynkin diagram for this space is

\be
 \begin{picture}(192,10)(0,0) \put(0,0){\makebox[0pt][l]{$\bt
 \hspace {2em}
 \ominus \hspace{2em}\times
 \hspace{2em}\ominus \hspace{2em}\bt$}
 \rule[.5ex]{11.5em}{.1ex} }
 \put(-5,10){ $n_1$ \hspace{1em} $n_2$
 \hspace {1.5em} $n_3$\hspace {1em} $n_4$
 \hspace {1em} $n_5$}
 \end{picture}\la{an2}
 \ee

An irreducible representation is specified by the Dynkin labels
$[n_1,\ldots n_5]$. Tensor fields will transform linearly under
the two $\gsl(2|1)$ subalgebras. A general $\ggl(2|1)$ Young
tableau is specified by three labels $<m_1,m_2,m_3>$ and these are
related to the (left-hand) $\gsl(2|1)$ Dynkin labels by

\be m_2-m_1=n_1\qquad m_2+m_3=n_2 \ee

so that the $\gsl(2|1)$ representation is left unchanged by adding
$m$ to $m_1,m_2$ and subtracting $m$ from $m_3$. For
representations with $m_1\neq 0$ it is possible to use this
freedom to bring the tableau to the canonical form, $m_1=0$. There
are two classes of representations, the long, or typical ones,

\be \setlength{\unitlength}{.2mm}
\begin{picture}(120,80)(60,60)
\put(5,120){$\cR_{ab}=$}
 \put(80,140){\framebox(100,20){$a$}}
\put(60,60){\framebox(20,100){$b$}}
\end{picture}
 \ee

corresponding to the $\gsl(2|1)$ labels $[a,a+b]$, and the short,
or atypical representations,

\be \cR_{a}= \overbrace{\square \! \square \cdots \square \!
\square}^{a+1} \ee

for which the labels are $[a,a+1]$. The latter representations
have non-negative, integral values for $a$, whereas the former
have non-negative integral values for $a$ but $b$ can be any real
number such that $b>1$.

If we let $b=1$ we apparently get $\cR_{a1}=\cR_{a}$. However,
this is not so. In fact, as one lets $b$ tend to 1, one finds that
the representation $R_{ab}$ becomes reducible at $b=1$, although
not completely reducible. We therefore have

\be \lim_{b\mapsto 1}\cR_{ab}\cong \cR_a + \cR_{a-1} \ee

This point was discussed explicitly in~\cite{Heslop:2001gp}.

Now let us consider the $\gg\gl(2|2)$ case relevant to $N=4$
superconformal symmetry. In this case there are again two types of
representation which we denote
 $\cR_{abc},\ \cR_{ac}$ with Young tableaux and Dynkin
 labels

 \be
 \setlength{\unitlength}{.2mm}
\begin{picture}(310,160)(0,0)
\put(0,80){$\cR_{abc}=$\ } \put(100,140){\framebox(180,20){$a$}}
\put(60,0){\framebox(20,80){$c$}} \put(60,80){\framebox(20,80){}}
 \put(80,80){\framebox(20,80){$b$}}
\put(120,60){$[a,a+b,c],\ b>1,\ a,c\geq0 $}
\end{picture}
 \ee

 \be
 \setlength{\unitlength}{.2mm}
\begin{picture}(300,100)(0,60)
\put(0,120){$\cR_{ac}=$} \put(80,140){\framebox(180,20){$a+1$}}
\put(60,60){\framebox(20,80){$c$}}
\put(60,140){\framebox(20,20){}}
\put(120,90){$\ba{ll} [a, a+1, c]& a,c\geq0\\
\left[0,0,c+1\right]&
  a=-1,c\geq0 \ea$}
\end{picture}
 \ee

There is also the trivial representation, $\cR_0$, which must be
treated separately in this case. For these representations $a$ and
$c$ are integral while $b$ can again be non-integral. Again we
observe that the limit of $\cR_{abc}$ as $b\mapsto 1$ appears to
be $R_{ac}$ but this is not so. In fact we find

\be
 \lim_{b\mapsto 1}\cR_{abc}\cong\cR_{ac} + \cR_{a-1,c+1}
 \ee

In terms of Dynkin labels this reads

 \bea
 \lim_{b\mapsto 1}[a,a+b,c]&\cong&[a,a+1,c]+[a-1,a,c+1]\qquad a\geq
 1\la{63}\\
 \lim_{b\mapsto 1}[0,b,c]&\cong&[0,1,c] + [0,0,c+2]
\la{64}
 \eea

As in the $N=2$ case one can carry out this limiting procedure
explicitly. The representations $\cR_{abc}$ for fixed $a$ and $c$
all have the same number of components for any real value of
$b>1$. When one takes the limit one finds explicitly that the
representation becomes reducible at $b=1$ according to the pattern
we have just described. This can be seen explicitly by taking
traces of the representations as was done in~\cite{Heslop:2002hp}
(although a different representation of the Young tableaux was
used in that paper).

In the field theory context the continuous label $b$ is related to
the anomalous dimension of an operator, and the limit $b\mapsto 1$
can be viewed as the classical limit. The reducible representation
obtained at the limit can be viewed as the ancestor of the
descendant representation given by the smaller of the two
representations (the second one) on the right-hand side of
equation \eq{63} or \eq{64}.

There are operators which transform under limiting representations
for both the left and right $\gg\gl(2|2)$s. In this case the
original representation will split into four in the limit. An
example of this is provided by the Konishi operator,
$[0(1+b)000(1+b)0]$ which splits into the irreducible
representations $[0100010]$, $[0100200]$, $[0020010]$ and
$[0020200]$ at $b=1$. These correspond to the free Konishi
operator, two series B descendants and a series C descendant. As
noted in \cite{Dolan:2002zh} series C operators of the form
$[001p100]$ cannot arise as descendants, as one can see from
\eq{64}, and so must be protected. It also  follows, from \eq{63},
that series A short representations of the form
$[a(a+1)0p0(a+1)a]$ cannot arise as descendants although they can
occur as limiting reducible representations. A similar statement
holds true for short series B representations of the form
$[a(a+1)qp(q+a+2)00]$.

\section{Two- and three-point functions}

In \cite{Heslop:2001gp} it was shown how one can solve the
$\psl(4|4)$ Ward identities and write down 2- and 3-point
functions of arbitrary protected operators in $N=4$ SYM. It was
also argued that these formulae can be extended to unprotected
operators, which can have non-integer dilation weight, if we
extend the definition of a tensor superfield to quasi-tensor
superfields. The formulae can also be generalised
straightforwardly to 4-point functions and higher n-point
functions in which case we also have to include functions of
invariants. It was also shown that all 2-, 3- and 4-point
functions of protected operators are automatically covariant under
the bonus $U_Y(1)$ symmetry. At first sight it appears that this
argument might extend to unprotected operators as well. This is
not the case as we shall see. In this section we shall only
consider representations with (half)-integer dilation weights. In
analytic superspace these can all be written as analytic tensor
superfields and the point about $U_Y(1)$ covariance can be
illustrated in this case.

We shall be considering solutions to the $\psl(4|4)$ Ward
identities for an $n$-point correlator,

\be <12\dots n>:= <\cO^{Q_1}_{\cR_1 \cR'_1}(X_1)\dots
\cO^{Q_n}_{\cR_n \cR'_n}(X_n)> \ee

 where the
operators transform as in~\eq{trans}. The Ward identities state
that the correlator must be invariant under superconformal
transformations. In other words

\be \d<12\dots n> = \sum_{i=1}^n (\cV_i + \cR_i(A_i) + \cR'_i(D_i)
+ Q_i \D_i)<12\dots n>=0 \la{ward} \ee

where $A_i:=A(X_i)\ D_i:=D(X_i)$. Throughout this section, we
shall assume that the Young tableaux corresponding to the
representations $\cR_i$ are in canonical form. As we have seen, it
is always possible to make this choice. A discussion of the
$U_Y(1)$ properties of correlation functions in Minkowski
superspace was givne in~\cite{Park:1999pd}.

\subsection{Two-point functions}

First we shall consider the 2-point formula given
in~\cite{Heslop:2001gp}:

\be
 <\cO^Q_{\cR\cR'}(1)\cO^Q_{\cR'\cR}(2)>\sim (g_{12})^Q
 \cR(X_{12}^{-1}) \cR'(X_{12}^{-1})
 \la{10}
 \ee

where $X_{12}:=X_1-X_2$, and the propagator $g_{12}$ is defined by

 \be
 g_{12}:={\rm sdet} X_{12}^{-1}={\hat y_{12}^2\over
 x_{12}^2}={y_{12}^2\over \hat x_{12}^2}
  \ee

where

 \bea
 \hat x_{12}&=& x_{12} - \l_{12} y_{12}^{-1} \p_{12}\\
 \hat y_{12}&=& y_{12} - \p_{12} x_{12}^{-1} \l_{12}
 \la{12}
 \eea

and matrix multiplication is implied, with the inverses having
downstairs indices $(x^{-1})_{\adt\a}$, $(y^{-1})_{a'a}$. In the
above formula, one takes $I$ factors of $X_{12}^{-1}$ ($I$=total
number of primed or unprimed indices), with the unprimed indices
in the $\cR$ representation (in other words, taking
symmetric/antisymmetric combinations as dictated by the
(canonical) Young tableau for the representation $\cR$), this
automatically puts the primed indices in the $\cR$ representation
as well. Similarly, one then takes another $I$ factors of
$X_{12}^{-1}$, with the primed indices in the $\cR'$
representation (which also puts the unprimed indices in this
representation).

One can check that this satisfies the Ward identities~\eq{ward},
given that

\bea
\cV X_{12}& = &A_1 X_{12} + X_{12} D_2 \nn\\
           & = &A_2 X_{12} + X_{12} D_1 \la{X}\\
\cV g_{12}&=&-(\D_1 +\D_2)g_{12}.\la{prop} \eea

The propagator takes care of the $Q_i \D_i$ terms, and the $Xs$
take care of
 the $\gsl(2|2)$ transformations.

In this formula for the 2-point function we have assumed that the
Young tableau for $\cR$ is in canonical form \eq{can}. We know,
however, that there are other Young tableaux which can specify the
same $SL(2|2)$ representation~\eq{sim}. We can ask whether they
might not be used instead to satisfy the Ward identities. This is
indeed true, although in the case of the 2-point function we
obtain the same answer as before so the 2-point function is
unique. For higher-point functions we can obtain different
solutions in this way.

To illustrate this point, consider the replacement of the Young
tableau $\cR=<0,n_1,n_2-n_1,n_3+n_2-n_1>$ by  the Young tableau
$\cS=<m,n_1+m,n_2-n_1-m,n_3+n_2-n_1-m>$ instead of $\cR$ which
carries the same representation of $\gsl(2|2)$. One can show that

\be <\cO^Q_{\cR\cR'}(1)\cO^Q_{\cR'\cR}(2)>\sim
(g_{12})^{Q-m}\cE(m)^{-1} 
 \cS(X_{12}^{-1})\cE(m) \cR'(X_{12}^{-1})\la{11}
 \ee

is also a solution to the Ward identities. This is, however, a
trivial statement as this expression is in fact equal to the
previous expression in~\eq{10}. In fact, if two representation
spaces are related by $\cS=\cE(m)\cR$, then we have

 \be
 \cS(X_{12}^{-1})=(\sdet X_{12})^{-m}
\cE(m) \cR(X_{12})\cE(m)^{-1}=(g_{12})^m \cE(m)
\cR(X_{12})\cE(m)^{-1}
 \la{RO}
 \ee

which is the finite version of equation \eq{+m}, valid as long as
$X^{-1}_{12} \in GL(2|2)$. This formula is easy to check: it is
obviously true in the case $X \in SL(2|2)$, since both $\cR$ and
$\cS$ have the same Dynkin labels, and thus correspond to the same
$SL(2|2)$ representation. Clearly both sides of the equation are
representations of $GL(2|2)$, so we just have to show they are the
same representation. One therefore just needs to check that the
equation is true for the matrix $Y=\mbox{diag}(d,d|d^{-1},d^{-1})$
since one can obtain all $GL(2|2)$ matrices by multiplying an
$SL(2|2)$ matrix by $Y$. In fact one just needs to check that the
highest weight state transforms in the same way under $Y$, the
correct transformation properties of all the other states is
guaranteed at the Lie algebra level as we can generate all other
states by applying lowering operators from $\gs \gl(2|2)$.

\subsection{Three-point functions}

The formula for 3-point functions, given in~\cite{Heslop:2001gp},
is

\be \ba{rcl}
 <123>&
 \sim&(g_{12})^{Q_{12}}(g_{23})^{Q_{23}}
 (g_{31})^{Q_{31}}\qquad \xz \\
 \\ &&\cR_2(X_{12}^{-1})_{\unA_2 \unB'_2}\cR'_2
 (X_{12}^{-1})_{\unB_2 \unA'_2}
 \cR_3(X_{13}^{-1})_{\unA_3 \unB'_3}
 \cR'_3(X_{13}^{-1})_{\unB_3 \unA'_3}\xz
\ t(X_{123})^{\unB_2 \unB_3 \unB'_2 \unB'_3}_{\unA_1 \unA'_1}
 \la{17}
 \ea \ee

where

 \bea
 <123>&=& <\cO^{Q_1}_{\unA_1\unA'_1}(1)\cO^{Q_2}_{\unA_2\unA'_2}(2)
 \cO^{Q_3}_{\unA_3\unA'_3}(3)>\\
Q_{ij}&=&{1\over2}(Q_i + Q_j - Q_k),\qquad k\neq i,j
 \la{18}\\
 X_{123}&=&X_{12}X_{23}^{-1}X_{31}
 \la{19}
 \eea

and the notation $\unA_i \ i=1,2,3$ represents all the unprimed
indices of the operator at the point $i$, which together carry the
representation $\cR_i$, and similarly for $\unA'_i$. Then the
indices $\unB'_i \ i=2,3$ must also carry the representation
$\cR_i$ and $\unB_i$ must carry the representation $\cR'_i$.

The tensor $t$ is a monomial function of $(X_{123})^{AA'}$,
$(X^{-1}_{123})_{A'A}$, $\d_A^B$ and $\d_{A'}^{B'}$ with index
structure as shown. In general there are many different possible
such monomials and the complete three-point function will be a
linear combination of all the possibilities. We will get further
restrictions on the allowed non-vanishing correlation functions
due to analyticity in the internal $y$ coordinates which we will
consider later.

In contrast to the case of 2-point functions, we can obtain new
solutions to the Ward identities at 3 points  by using Young
tableaux which are not in canonical form. This results in
solutions of the same form as \eq{18} but with factors of $\cE$.
Because $\cE$ is not $\pgl(2|2)$ invariant, the powers of the
propagators have to be adjusted accordingly.

\subsection{Transformation of correlators under $PGL(4|4)$}

$N=4$ super Yang-Mills is only invariant under $PSL(4|4)$ not
$PGL(4|4)$, but we can ask what implications the invariance under
$PSL(4|4)$ has for $PGL(4|4)$ transformations. In particular, we
can examine the transformation of the correlators, obtained in the
previous section by solving the $PSL(4|4)$ Ward identities, under
the enlarged group $PGL(4|4)$.

In the examples to be discussed below we shall suppose that all of
the operators have canonical tableaux.  We shall say that such a
correlator has $U(1)_Y$ charge $p$ if

\be \sum_{i=1 \dots n} (\cV_i + \cR_i(A_i) + \cR'_i(D_i) + Q_i
\D_i)<1 \dots n>=p\D' <1\ldots n>\ee

where $\cV$ defines a $\pgl(4|4)$ transformation.  Note that this
definition makes sense even if one cannot assign a definite
$U(1)_Y$ charge to a given operator.

The following two results can be seen relatively
straightforwardly:

$\bullet$ All 2-point functions have vanishing $U_Y(1)$ charge.

To see this simply consider the expression for the 2-point
function $<12>$~\eq{10}. Since $\cR$ and $\cR'$ are Young tableaux
such that $m_1=m'_1=0$, one finds, under a $\pgl(4|4)$
transformation,  that

\be (\cV_1 +\cV_2 + \cR(A_1) + \cR'(A_2) + \cR'(D_1)+\cR(D_2) +
Q(\D_1+\D_2))<12>=0 \ee

where $\cV$ is the vector field generating this transformation.
This follows because the formulae given in \eq{X} and \eq{prop}
remain valid in $\pgl(4|4)$, as one can easily verify. Thus the
2-point function has vanishing $U_Y(1)$ charge since the Young
tableaux are in the canonical form with $m_1=m'_1=0$ and there is
no term involving the $U_Y(1)$ parameter $\D'$.

Note that this result does not necessarily mean that the 2-point
function is $PGL(4|4)$ invariant. If one considers  a two-point
correlator of the form of an operator and its conjugate, the total
$U(1)_Y$ charge is zero implying $PGL(4|4)$ invariance. However,
one can also consider the two-point function of one operator and
the conjugate of another which transforms in the same way under
$PSL(4|4)$ but which has different $U(1)_Y$ charge. Since $U(1)_Y$
is not a symmetry of the interacting theory, there is no reason
why such correlators should vanish. In this case, the $U(1)_Y$
charge of the correlator would still be zero whereas the charge
required by $PGL(4|4)$ symmetry would be equal to that of the
charged operator. This applies particularly to long operators. One
would therefore expect the diagonal combinations not to have
well-defined $U(1)_Y$ charges. In principle,  the same reasoning
should apply to semi-protected operators.

$\bullet$ All 2-,  3- and 4-point functions of protected operators
have vanishing $U_Y(1)$ charge and are invariant under $PGL(4|4)$.
This follows because the Young tableaux for protected operators
have $m_1=m'_1=0$ as mentioned at the end of section~\ref{sec:YT}.
For $n=2,3$ the $n$-point functions satisfy the equation

\be \sum_{i=1 \dots n} (\cV_i + \cR_i(A_i) + \cR'_i(D_i) + Q_i
\D_i)<1 \dots n>=0 \ee

where again $\cV$ is the vector field generating the $\pgl(4|4)$
transformation. Since all Young tableaux in this expression must
be in the canonical form (i.e. with $m_1=0$) this has charge zero.
In this case the correlator is $PGL(4|4)$ invariant (i.e.
satisfies $PGL(4|4)$ Ward identities $\d<1\dots n>=0$.) This is
because there are no protected operators with non-zero $U(1)_Y$
charge.

For four or more points, such correlators can be written in terms
of explicit functions involving the $Xs$ and the propagators
$g_{ij}$ multiplied by functions of the invariants. Because there
is no freedom in the choice of tableaux for such operators, the
former give rise to $U(1)_Y$ invariant expressions, and so the
$U(1)_Y$ invariance of $n$-point functions of protected operators
depends on the $U(1)_Y$ properties of the invariants. As we know,
the 4-point invariants are invariant under $U(1)_Y$ while this is
not necessarily true at 5 or more points~\cite{ehw}. This will be
discussed in more detail in section 7.

\subsection{$<L A_{Q_2} A_{Q_3}>$}

We denote the chiral primary operators by $A_Q,\ A_Q:=\tr (W^Q)$.
A 3-point function with one arbitrary operator and two chiral
primary operators will have vanishing $U_Y(1)$ charge as predicted
in~\cite{Intriligator:1999ff}. From the formula for three point
functions~\eq{17} we obtain

\be <\cO^{Q_1}_{\cR \cR} A_{Q_2} A_{Q_3}> \sim \cR(X_{123}^{-1})
g_{12}^{Q_{12}} g_{13}^{Q_{13}} g_{23}^{Q_{23}} \ee

We can again assume that $\cR$ is in canonical form, and thus we
can see that this expression has zero $U_Y(1)$ charge. As in the
case of two-point functions this does not always mean the
correlator is $PGL(4|4)$ invariant, since one could have an
operator $\cO_{\cR \cR'}^{Q,Q'}$ on the left-hand side with
non-zero $U_Y(1)$ charge, but the right-hand side would still have
zero $U_Y(1)$ charge.

\subsection{A correlation function with non-zero $U_Y(1)$ charge}

We shall now consider examples of correlation functions which do
have non-zero $U_Y(1)$ charge. The simplest example one can
construct involves two series C vector operators and one long
operator

 \be
 <123>=<\cO_{\unA \unA'}^{Q_1} \cO_{BB'}^{Q_2} \cO_{CC'}^{Q_3}>
 \ee

where both sets of indices on the first operator are in the
representation $\cR=<0033>$, $Q_1=6+d,\ Q_2=2 + d_2,\ Q_3=2+d_3$.
The $d$s can be identified with the central Dynkin label of the
representations. The operators at points 2 and 3 are such series C
operators with internal quantum numbers given by $[1 d_i 1],\
i=2,3$.

The  solution \eq{18} to the ($\psl(4|4)$) Ward identities is

 \be
 <123>\sim g_{12}^{Q_{12}} g_{13}^{Q_{13}} g_{23}^{Q_{23}}
 (X_{12}^{-1})_{B'D} (X_{12}^{-1})_{D'B}
 (X_{13}^{-1})_{C'E} (X_{13}^{-1})_{E'C}\,
 t^{DD',EE'}_{\unA\unA'}(X_{123})
 \ee

There are many possible solutions depending on the choice of $t$.
One possibility is

 \be
 t^{DD',EE'}_{\unA\unA'}=
 P\left(\d_{A_1}^{D}\d_{A_2}^{E}\d_{A'_1}^{D'}\d_{A'_2}^{E'}
 \cS(X_{123}^{-1})_{\hat{\unA}\hat{\unA'}}\right)
 \ee

where we have written $\unA=(A_1,A_2,\hat{\unA})$ and similarly
for the primed indices. The representation $\cS$ corresponds to
the tableau $<0022>$ and $P$ projects the $\unA$ and $\unA'$
indices onto the tableau $<0033>$. One can check that this is
analytic in the internal coordinates if

\bea
 d + d_2 -d_3 &\geq & 2 \\
 d + d_3 -d_2 & \geq & 2 \\
 d_2 + d_3 -d & \geq & 4.
\eea

Under a $\pgl(4|4)$ transformation this solution satisfies the
equation

\be \sum_{i=1 \dots 3} (\cV_i + \cR_i(A_i) + \cR'_i(D_i) +Q_i
\D_i) <123> =0 \ee

and therefore has vanishing $U_Y(1)$ charge.

However, we can also exhibit a solution involving the
$\cE$-tensor. It is

 \be
 <123>\sim g_{12}^{Q_{12}-\fft12}
g_{13}^{Q_{13}-\fft12} g_{23}^{Q_{23}+\fft12}
 (X_{12}^{-1})_{B'D} (X_{12}^{-1})_{D'B}
 (X_{13}^{-1})_{C'E} (X_{13}^{-1})_{E'C}
 \tilde t^{DD',EE'}_{\unA\unA'}(X_{123})
 \ee

where

 \be
 \tilde t^{DD',EE'}_{\unA\unA'}=
 \cE^{-1}\tilde P\left(\d_{A_1}^{D}\d_{A_2}^{E}\d_{A'_1}^{D'}\d_{A'2}^{E'}
 \cS(X_{123}^{-1})_{\hat{\unA}\hat{\unA'}}\right)
 \ee

where $\tilde P$ projects onto $<0033>$ on the unprimed covariant
indices and onto $<1122>$ on the primed indices. The $\cE^{-1}$
then acts on the primed indices to bring them back into the
$<0033>$ pattern.

This solution is valid (i.e. analytic in the internal coordinates)
when

\bea
 d + d_2 -d_3 &\geq &1 \\
 d + d_3 -d_2 & \geq & 1 \\
 d_2 + d_3 -d & \geq & 1.
\eea

Under a $PGL(4|4)$ transformation this solution satisfies

\be \sum_{i=1 \dots 3} (\cV_i + \cR_i(A_i) + \cR'_i(D_i) +Q_i \D_i
)<123>= \D' <123> \ee

This is again a solution of the superconformal Ward identities
(since for $PSL(4|4)$ transformations $\D'=0$) but it has
non-vanishing $U_Y(1)$ charge.

One would expect a general 3-point function of this type to be
given by a linear combination of the above solutions (and other
possible solutions) so that the 3-point function of such operators
will not have a well-defined $U_Y(1)$ charge.

This example can be generalised to more complicated three-point
functions of the same type, i.e one-quarter BPS series C operators
at points 2 and 3 and a long operator at point 1 with suitably
chosen representations. The construction can be further extended
in two ways: the long operator can be allowed to be a quasi-tensor
(which is the interesting case in practice), and it can be
replaced by a semi-protected operator. In all of these cases the
basic principle is the same: one can find two (or more) different
solutions of the Ward Identities involving two (or more) different
representations of $\ggl(2|2)$ which are equivalent in
$\gsl(2|2)$.

\subsection{Semi-protected operators}

The proof that 3- and 4-point functions involving protected
operators are $U_Y(1)$ invariant depended on both the left and
right $GL(2|2)$ representations of both operators being short and
therefore does not apply to series B operators which do not
saturate the series B bound or to series A operators which only
saturate one unitary bound. This is because for these operators,
although one $GL(2|2)$ representation is short, the other is long.
We call such operators semi-protected; even though the operators
themselves are non-renormalised~\cite{Heslop:2001dr}, 2- and
3-point functions involving them are in general not $U_Y(1)$
invariant. Despite this, however, it is still possible to show
that the 2-point functions of semi-protected operators are
non-renormalised using a slight generalisation of the usual
argument.

The non-renormalisation of the two-point functions of
semi-protected operators is proved using the reduction formula
which relates the derivative of the two-point function with
respect to the complex coupling $\t$ to the three point function
involving an insertion of the energy-momentum supercurrent:

\be {\del \over \del \t} <\cO \bar \cO>\sim \int {\rm d}\m_0 <T_0
\cO \bar \cO>\la{red} \ee

where ${\rm d}\m_0={\rm d}^4x{\rm d}^4\l{\rm d}^4y$. It will be
important to note that the measure has $U_Y(1)$ charge +2.

We now show that the correlation function involving $T$, one
semi-protected operator, $\cO$, and its conjugate operator,
$\bar\cO$,

\be <T  \cO^{Q}_{\unA \unB'} \bar \cO^{Q}_{\unB \unA'}>\label{ABB}
\ee

can have $U_Y(1)$ charge at most 1 so the right-hand side
of~\eq{red} vanishes. Note that previous non-renormalisation
theorems relied on the fact that the correlation functions were
$U_Y(1)$ invariant, i.e. had charge zero. Here the indices $\unA$
and $\unA'$ are in the same short representation $\cR$ which has
an L-shaped Young tableau $<0,m_2,1,m_4>$, while the $\unB$ and
$\unB'$ indices are in an arbitrary (long) representation $\cS$
which we assume to be canonical ($m_1=0$).

Before writing down the three-point function, we make a small
digression on the three-point function of two vector operators and
one $T$. From the general formula \eq{17} we have

 \be
 <T\cO_{AB'}\cO_{BA'}>\sim(X_{12}^{-1})_{C'A}(X_{12}^{-1})_{B'D}
 (X_{13}^{-1})_{D'B}(X_{13}^{-1})_{A'C}t(X_{123})^{CC',DD'}
 \ee

where we have omitted the propagator factors. There are two
possible choices for $t$:

 \bea
 (a)\qquad t^{CC',DD'}&=&(X_{123})^{CC'}(X_{123})^{DD'} \\
 (b)\qquad t^{CC',DD'}&=&(X_{123})^{CD'}(X_{123})^{DC'}
 \eea

These lead to the following solutions for the three-point
function:

 \bea
 (a)\qquad
 <T\cO_{AB'}\cO_{BA'}>&\sim&(X_{23}^{-1})_{A'A}(X_{23}^{-1})_{B'B}\\
 (b)\qquad<T\cO_{AB'}\cO_{BA'}>&\sim&(X_{312}^{-1})_{A'B}(X_{231}^{-1})_{B'A}
 \eea

The idea now is to rewrite the three-point function \eq{ABB}
making use of these two basic solutions. That is, we factorise the
$\cR$ and $\cS$ representations, and write a solution of type (a)
for the first factor multiplied by a solution of type (b) for the
second factor. In addition, in order to obtain a non-zero $U(1)_Y$
charge, we shall have to change the primed (or unprimed) long
representation $\cS$ to a non-canonical one, $\tilde\cS$, with
$m'_1=m$, say. We then have to supply an appropriate factor of
$\cE$ to return the indices into the same representations as the
left-hand side. A solution of this form is

\be \ba{l}
 <T
\cO^{Q}_{\unA \unB'} \bar \cO^{Q}_{\unB \unA'}> \sim
g_{12}^{\phantom Q} g_{13}^{\phantom Q} g_{23}^{Q-1-m}\\ \\
\times \cE(m)^{-1}\,P\left( \cR_1(X_{23}^{-1})_{\unA_1 \unA'_1} \
\cR_2(X_{23}^{-1})_{\unB_1\unB'_1}\
 \cR_3(X_{213}^{-1})_{\unB_2 \unA'_2} \
 \cR_4(X_{312}^{-1})_{\unA_2 \unB'_2}\right).
\ea \ee

where we have split the multi-indices into two sets
$\unA\rightarrow (\unA_1,\unA_2)$ etc. The operator $P$ projects
the $\unA$ and $\unA'$ indices onto the representation $\cR$, the
$\unB$ indices onto the representation $\cS$ and the $\unB'$
indices onto the $\tilde\cS$ representation. Finally, the $\cE$
tensor acts on the $\unB'$ indices to bring them back into the
representation $\cS$.  The $\tilde\cS$ Young tableau is in
non-canonical form with $m'_1=m$. It defines the same $SL(2|2)$
representation as $\cS$ but they differ in $GL(2|2)$. Such a
solution, if it exists, will have $U(1)_Y$ charge $m$.

For the projection operator $P$ to give a non-zero result we
clearly require that $\cR$ is contained in the tensor product
$\cR_1 \otimes \cR_3$, and similarly for the other indices. So we
must have

\bea
\cR_1 \otimes \cR_3 &\ni& \cR\label{r1}\\
\cR_1 \otimes \cR_4 &\ni& \cR\label{r2}\\
\cR_2 \otimes\cR_3 &\ni& \cS\label{r3}\\
\cR_2 \otimes \cR_4 &\ni& \tilde\cS\label{r4}. \eea

Now clearly a long representation multiplied by any other
representation will still be long. Therefore, since $\cR$ is short
\eq{r1} and \eq{r2} imply that $\cR_1$, $\cR_3$ and $\cR_4$ are
all short representations. Similarly, a representation in
canonical form (with $m_1=0$) cannot be obtained in the  tensor
product of a representation with $m_1 \neq 0$ and any other
representation (this can be seen by considering the multiplication
of Young tableaux). Since $\cS$ is in canonical form, by
assumption,~\eq{r3} implies that
 $\cR_2$ must be in canonical form
($m_1=0$). Finally, \eq{r4} tells us that $\tilde\cS$ is contained
in the tensor product of a Young tableau in canonical form
($\cR_2$) and a short representation($\cR_4$).

For a semi-protected operator in series A, the Young tableau for
$\cR$ has the form $<0,k,1,q+2r+s-(k+1)>$ while the Young tableau
for $\cS$ has the form $<0,s,r,r+q>$. Now let us suppose that
$\tilde\cS$ is $<2,s+2,r-2,(r+q)-2>$, so that $m=2$, and that the
tableau for $\cR_2$ is $<0,p_2,p_3,p_4>$. When $\cR_2$ is
multiplied by any representation, the third and fourth tableau
numbers cannot be diminished, so that, at most, $p_3=r-2$ and
$p_4=(r+q)-2$. Suppose this is the case, then since the two
left-most columns of $\cS$ and $\tilde\cS$ differ by a subtableau
in the form of a $2\xz 2$ square, it follows that the $\cR_3$
factor of the short representation $\cR$ should contain such a
subtableau. But this is impossible since $\cR$ is a short tableau
in the form of an $L$ (with arms which both have single-box
width). Clearly the situation is not improved by reducing $p_3$
and/or $p_4$, so we conclude that there is no solution with $m=2$.
The situation is even more constrained as we increase $m$, whereas
it is possible to obtain $m=1$. We therefore conclude that the
maximum value of the $U(1)_Y$ charge of the three-point correlator
\eq{ABB} is $1$. This holds in series A, but it is also true in
series B. In fact, the result is easier to see in this case
because $\cR$ only has $m_4\neq 0$, and this certainly cannot have
a subtableau in the form of a $2\xz 2$ square.

In conjunction with the reduction formula this result implies that
two-point functions of semi-protected operators are
non-renormalised since the measure on the right-hand side of the
reduction formula has $U(1)_Y$ charge $-2$.

Such arguments do not seem to extend to the case of three-point
functions, however, and we expect three-point functions involving
semi-protected operators to receive quantum corrections in
general. Indeed, as we have remarked, three-point functions with
one semi-protected operator and two series C operators will in
general violate $U(1)_Y$ symmetry.

\subsection{Comparison with earlier results}

In reference \cite{Intriligator:1999ff} it was conjectured that
three-point functions of short operators and three-point functions
with one long and two short operators should obey the $U(1)_Y$
selection rule, i.e. that such correlators would vanish unless the
sum of the $U(1)_Y$ charges is zero. As we have seen, in the
interacting theory, it is not clear that a general long or
semi-protected operator has a well-defined $U(1)_Y$ charge, but we
can nevertheless rephrase the conjecture in terms of the $U(1)_Y$
charges of the correlators which we have defined for the case when
all the operators involved have canonical Young tableaux. In
addition, short operators were taken to be CPOs in
\cite{Intriligator:1999ff}, i.e. single-trace one-half BPS
operators, and long operators were also taken to be single-trace.
We can therefore update these earlier results by replacing the
short operators of \cite{Intriligator:1999ff} by arbitrary
protected operators, and by allowing semi-protected operators and
arbitrary long operators.

For the two-point functions we agree with
\cite{Intriligator:1999ff} in the sense that, as we have seen, the
$U(1)_Y$ charge of any  two-point function must vanish. However,
this does not necessarily imply $U(1)_Y$ symmetry of such
correlators because of the fact that this group is not a symmetry
of the interacting theory. As we have mentioned, one can therefore
have non-vanishing correlators of two operators which transform
under conjugate representations of $\psl(4|4)$ but which have
different $U(1)_Y$ charges. The diagonalisation of sets of long or
semi-protected operators with the same $\psl(4|4)$ quantum numbers
would therefore be expected to yield orthogonal operators with
ill-defined $U(1)_Y$ charges.

For three-point functions of short operators we agree with
\cite{Intriligator:1999ff}. Indeed, all protected operators have
zero $U(1)_Y$ charges (as superfields; the higher components
within a supermultiplet will be charged), and all two- and
three-point functions of such operators also have vanishing
$U(1)_Y$ charge and are invariant under $\pgl(4|4)$
\cite{Heslop:2001gp}.

Three-point functions with two CPOs or, more generally, two
one-half BPS operators and one long or semi-protected operator
will have vanishing $U(1)_Y$ charge as we have seen, although this
does not necessarily imply $U(1)_Y$ invariance. This is in
agreement with \cite{Intriligator:1999ff} in the sense that the
charge of the correlator vanishes. On the other hand, three-point
functions of two more general protected operators and one long or
semi-protected operator need not have vanishing $U(1)_Y$ charge.
As we have shown, there are three-point functions with two series
C one-quarter BPS operators for which one can explicitly exhibit
solutions with zero and non-zero $U(1)_Y$ charges indicating that
the charge of the correlator itself is not well-defined.

\section{Superconformal invariants in analytic superspace}

In this section we shall consider the construction of rational
superconformal invariants (under the group $PSL(4|4)$) of $n$
points in analytic superspace. These form a ring $\cI$ which has a
nilpotent ideal  $\cN$. All of the elements of the quotient ring
$\cQ:={\cI/\cN}$ were given in \cite{Howe:1996rk}, but it has
subsequently become apparent that there are indeed nilpotent
invariants \cite{ehw} starting at $n=5$ points. In fact, as
pointed out in \cite{Intriligator:1998ig}, elements of $\cQ$ are
invariant under $U(1)_Y$ and hence under the larger group
$PGL(4|4)$. The fact that there are no nilpotent invariants for
$n\leq 4$ is one way of interpreting the non-renormalisation
theorems for 2- and 3-point correlators of protected operators. In
reference \cite{Howe:2001je} a sketch was given of how one might
construct the invariants completely. Here, we complete the
procedure, at least in principle, and also show how to detect the
presence of non-$U(1)_Y$ invariants in a straightforward way.

\subsection{Coordinate approach}

Consider the transformation

\be \d X_i=B + A X_i + X_i D + X_i C X_i,\ i=1,\ldots n,\ {\rm no\
sum\ on\ i} \ee

where $X_i$ is the supercoordinate $X_i^{AA'}$ of the $i$th point.
We are looking for functions $F(X_1,\ldots X_n)$ which are
invariant under the above. We first solve for translations $B$. If
we change coordinates to $(X_1,X_{1i}),i=2\ldots n$ we find that
$F$ is independent of $X_1$. Now consider the transformation of
$X_{1i}$ under $C$,

\bea
 \d_C X_{1i}&=&X_1C X_1-X_i C X_i \nn\\
 &\phantom{=}&-X_{1i}CX_{1i}+X_1CX_{1i}+ X_{1i} C X_1
 \eea

For the inverse we therefore have

\be \d_C X_{1i}^{-1}=C -(C X_1)X_{1i}^{-1}- X_{1i}^{-1}(X_1 C) \ee

At this stage we can regard $F$ as being a function of the $n-1$
inverses $X_{1i}^{-1}$ and change variables to $X_{12}^{-1}$ and
$n-2$ variables $Y_i$ defined by

\be Y_i^{-1}:= X_{1i}^{-1}-X_{12}^{-1},\qquad i=3,\ldots n \ee

We note, for future reference, that

\be Y_i=X_{12} X_{2i}^{-1} X_{i1}:= X_{12i} \ee

We have

\be \d_C Y_i=Y_i(C X_1)+(X_1 C)Y_i \ee

and so the invariance of $F$ under $C$ implies

\bea \d_C F&=&\left(C-(CX_1)X_{12}^{-1}-X_{12}^{-1}(X_1 C)\right)
{\del F\over\del X_{12}^{-1}} \nn \\
& \phantom{=}& + \sum_{i=3}^{n}\left(Y_i(C X_1)+(X_1
C)Y_i\right){\del F\over \del
 Y_i}\ \ =\ \ 0
 \eea

Now $F$ is independent of $X_1$ so the above should be valid for
arbitrary values of this coordinate. Taking $X_1=0$ we see that
$F$ does not depend on $X_{12}^{-1}$. Thus $F$ depends only on the
$(n-2)$ $Y_i$'s and the residual transformation reduces to linear
transformations of type $A$ and $D$. Hence, if $F$, as a function
of the $Y_i$,  is invariant under the linear $A$ and $D$
symmetries it will automatically be completely invariant.

The $A$ and $D$ transformations are

\be \d Y_i= AY_i + Y_i D \ee

so if we change variables again to $Y_3$ and $Z_i:=Y_i Y_3^{-1},\
i=4,\ldots n$, we find that the $Z$s do not transform under $D$,
only under $A$:

\be \d Z_i =AZ_i -Z_i A \ee

The $Z_i$s can also be written as

\be Z_i:=X_{12}\left(X_{2i}^{-1} X_{i1} X_{i3}^{-1}
X_{32}\right)X_{12}^{-1} \la{Z} \ee

The function $F(Y_3, Z_i)$ will then be an invariant if

\be \left((AY_3 + Y_3 D){\del\over\del Y_3} + \sum_{i=1}^n
(AZ_i-Z_i A){\del\over\del Z_i}\right)F=0 \ee

Now in $N=2$ analytic superspace we can easily solve for
invariance under $D$ transformations. This superspace is defined
in a similar way to $N=4$ analytic superspace but now the internal
indices $a,a'$ only take on one value each. The matrices $A$ and
$D$ are $\ggl(2|1)$ matrices with $\str A=\str D$. This means we
can take $A$ to be straceless and use $D$ to conclude that $F$
must be independent of $Y_3$. So invariants in $N=2$ analytic
superspace are functions of $(n-3)$ variables $Z_i$ which are
$(2|1)\xz (2|1)$ supermatrices and which transform under the
adjoint representation of $\gsl(2|1)$.

In $N=4$ the situation is a little more complicated because we
cannot remove the straces from the matrices $A$ and $D$ in an
invariant manner. However, a similar result can be obtained after
some redefinitions. By using all of the parameters in $D$ except
for the supertrace, one can show that $F$ depends only on the $Z$s
and $W:=\str Y_3$. The invariance condition is now

\be \sum_i [A,Z_i] {\del F\over\del Z_i} + \str (A+D) W{\del
F\over\del W}=0 \ee

If we replace $Z_i$ by

\be Z'_i:=MZ_i M^{-1}\la{redef} \ee

 where

\be M=\left(\ba{cc} W^{-{1\over 8}}& 0\\ 0&
W^{{1\over8}}\ea\right) \ee

and where each entry in $M$ is a $2\xz 2$ matrix,  we find

\be \d Z_i'=[A',Z'_i] \ee

Writing the parameter matrix in $2\xz 2$ block form as

\be A=\left(\ba{cc} A_0& \C\\ \Upsilon & A_1\ea\right) \ee

where $A_0,A_1$ are even and $\C,\D$ are odd, we find

\be A'=\left(\ba{cc} A'_0& \C'\\ \Upsilon'& A'_1\ea\right) \ee

where

\bea
A'_0&=& A_0-{1\over8}\str(A+D) \nn\\
A'_1&=& A_1+{1\over8}\str(A+D) \nn\\
\C'&=&W^{-\qu}\C\nn\\
\Upsilon'&=&W^{+\qu}\Upsilon \eea

Note that $\str A'=\half\str (A-D)=\D'$, the $U(1)_Y$ parameter.
So we can regard $A'$ as a general $\ggl(2|2)$ matrix. If we now
drop the primes on both the $Z$s and $A$ we find that an $n$-point
analytic superspace invariant $F$ is a function of $n-3$ $(2|2)\xz
(2|2)$ supermatrix variables $Z_i$ subject to the differential
constraint

\be \sum_{i=3}^n [A,Z_i] {\del F\over\del Z_i}=0 \la{inv} \ee

The finite version of the transformation of the $Z$s is

\be Z_i\mapsto GZ_i G^{-1} \la{invfin} \ee

where $G$ can now be regarded as an element of $GL(4|4)$.

Note that the unit matrix in $A$ drops out of \eq{inv}, so that
this equation expresses invariance under the adjoint action of
$\pgl(2|2)$ on the $Z$ variables. Furthermore, if we restrict $A$
to satisfy $\str A=0$, then $F$ will be invariant under
$PSL(4|4)$; however, if $F$ is invariant under unrestricted $A$
transformations then it will be invariant under $PGL(4|4)$. So
this approach gives a very simple way of distinguishing the two
types of invariant. Alternatively, if we use the finite version,
then \eq{invfin} defines a $PGL(2|2)$ transformation; if $F$ is
invariant under transformations of the $Z$s for which $\sdet G=1$
it will be $PSL(4|4)$ invariant, whereas if $G$ is unrestricted it
will be invariant under $PGL(4|4)$.

The final step in the construction of invariants is to solve
equation \eq{inv} (or \eq{invfin}). One way of doing this is to
look for invariant polynomials in the components of the $Z$s. For
a $k$th degree moynomial this is equivalent to looking for
invariant tensors of $k$ covariant and $k$ contravariant
$\bbC^{2|2}$ vectors. The simplest such invariants are given by
taking supertraces of the $Z$s and these coincide with the
invariants found in~\cite{Howe:1996rk}. However, for a
sufficiently large number of indices, one can use the $\cE$-tensor
which is invariant under $\gsl(2|2)$ but not under $\ggl(2|2)$ -
these invariants will clearly not be invariant under $U(1)_Y$.
This tensor has twelve indices and so requires a sixth degree
monomial in the $Z$s. The six covariant indices should be
projected onto the tableau $<0033>$, while the upper indices
should be put in the $<1122>$ tableau arrangement. Clearly, this
cannot be done with only one $Z$, and so the lowest number of
points at which one can construct a non-$U(1)_Y$ invariant is
therefore $n=5$ as we know from other arguments.

In summary, approaching the problem in this way, the $n$-point
invariants are formed from homogeneous polynomials of the $(n-3)$
$Z$ variables by suitably contracting the indices with $\d$s or
the invariant tensor $\cE$. Those which involve
the latter will not be invariant under $U(1)_Y$.

An alternative way to carry out the last step is to use up the
remaining parameters to reduce the number of components contained
in the $Z$s. Let us consider first the simpler case of $n=4$
points in ordinary Minkowski space. We can apply a very similar
analysis to this case and conclude that the $n$-point invariants
are given by $(n-3)$ $2\xz 2$ matrix variables $z_i$ which
transform at the final step by $z_i\mapsto gz_i g^{-1}$ where
$g\in SL(2)$. For $n=4$ we only have one $z$ variable, and so one
way of constructing the invariants would be to form traces of
products of $z$s. Only two of these are functionally independent,
so that there are two 4-point invariants as expected. Another
approach is to use the $g$ transformation to bring $z$ to a
diagonal form, $z=\diag (x_1,x_2)$. This form is invariant under
infinitesimal $A$-transformations for which $A$ is diagonal and
traceless. So the two invariants can be replaced by the
eigenvalues $x_1,x_2$. For more than four points we can use the
residual  $g$ transformation to remove one more variable leaving
us with $4n-15$ independent invariants for $n\geq 5$. We remark
that the variables $x_1,x_2$ are those used  in
ref~\cite{Dolan:2000ut}
 in their
discussion of 4-point functions (and denoted $x,z$ by the
authors).

Now let us return to the supersymmetric case. Again we start with
$n=4$ points, and only  one $Z$. By a finite transformation of the
form $Z\mapsto GZG^{-1},\ G\in GL(2|2)$ we can bring $Z$ to a
block-diagonal form. In other words, we can eliminate all of the
odd variables in $Z$. Moreover, it is in principle possible to
find the matrix $G$ which effects this change. We can then
diagonalise the two $2\xz 2$ even submatrices to make $Z$
diagonal, $Z=\diag (X_1,X_2|Y_1,Y_2)$. This means that there are 4
independent superinvariants at $n=4$ points, and they are all
invariant under $U(1)_Y$. They can also be written as $\str
(Z^p),\ p=1,\ldots 4$. At 5 or more points, we can eliminate the
odd variables of one $Z$, $Z_4$ say, as before. There is a
residual (infinitesimal) symmetry consisting of diagonal matrices
$A$ which  corresponds to only 3 parameters as the unit matrix
does not act. We can use two of these parameters to bring the even
$2\xz 2$ submatrices of $Z_5$, say, to triangular form, and this
form is preserved by $A$s of the form $A=\diag(aI_2, bI_2)$. So
the only residual transformation is in fact $U(1)_Y$ which acts
only on the odd coordinates, half with positive charge and half
with negative charge. At this point it looks as if we can
construct odd invariants as well as even ones, but these are in
fact not rational because of the fractional powers of $W$ used
above in the redefinitions of the $Z$s. The
redefinition~\eq{redef} affects only the odd coordinates and
requires them to occur in particular combinations. In fact, if we
write

\be Z=\left( \ba{cc} z& \th\\ \f & w\ea \right) \ee

then each invariant must involve the odd variables in the form
$\th^p\f^q$ where $p-q=0$ mod $4$. This is another way of spotting
$U(1)_Y$ invariance, since those invariants which do have this
symmetry will have $p=q$ in each contributing term.

\subsection{Grassmannian approach}

In reference \cite{Howe:1996rk} $N=4$ superconformal invariants in
analytic superspace were discussed in a slightly different
formalism. We briefly review this and then use the $\cE$ tensor to
construct the missing invariants, not invariant under $PGL(4|4)$.
We first consider the space $\bbU$ of $(2|2)\xz (4|4)$ matrices of
maximal rank. This space is acted on in a natural way by $GL(2|2)$
from the left and $GL(4|4)$ from the right. We can then form the
quotient space $GL(2|2)\bsh \bbU$. This can easily be identified
with the Grassmannian of $(2|2)$ planes in $\bbC^{4|4}$, and is
another description of analytic superspace. It is straightforward
to see that only $PGL(4|4)$ acts on analytic superspace.

We shall now consider the construction of $n$-point invariants on
this space. With each point $i$ we associate $u_i\in \bbU$. An
invariant $F$ is then a function of the $u$s which is invariant
under each $GL(2|2)_i$ (one for every point) separately, and under
$u_i\mapsto u_i g,\ g\in PGL(4|4)$ jointly. We can single out two
points, 1 and 2, say, and form the $(4|4) \xz (4|4)$ matrix
$u_{12}$ formed by  taking the upper rows to be given by $u_1$ and
the lower rows to be given by $u_2$. This matrix transforms from
the left by $h_{12}:=\diag (h_1,h_2),\ h_i\in GL(2|2)_i$. We now
define two families of $(2|2) \xz (2|2)$ matrices by

 \bea
 K_i:=u_i(u_{12}^{-1})^1,\qquad i=3,\ldots n\nn\\
 L_i:=u_i(u_{12}^{-1})^2, \qquad i=3,\ldots n\nn
 \eea

where the numerical superscripts on the inverse matrix denote the
projections onto the corresponding submatrices. These matrices are
invariant under $GL(4|4)$ and transform by

 \bea
 K_i \mapsto h_i K_i h_1^{-1}\nn\\
 L_i \mapsto h_i K_i h_2^{-1}\nn
 \eea

Set

 \be
 M_i=(K_i)^{-1} L_i,\qquad i=3,\ldots n
 \ee

to obtain $(n-2)$ matrices which transform only under $h_1$ and
$h_2$,

 \be
 M_i\mapsto h_1 M_i h_2^{-1}
 \ee

The matrices

 \be
 N_i=M_i M_3^{-1},\qquad i=4,\ldots n
 \ee

are invariant under $h_2$ and transform under the adjoint
representation of $GL(2|2)_1$,

 \be
 N_i\mapsto h_1 N_i h_1^{-1}
 \ee

$GL(2|2)$ invariants constructed from the $n-4$ $N$s will then be
invariant under $PGL(4|4)$. These invariants are the supertraces,
or, to put it another way, one can consider a polynomial in powers
of the matrix elements of the $N$s and hook the indices together
with the invariant matrix $\d$. These were the invariants
identified in~\cite{Howe:1996rk}. They are non-nilpotent and in
fact determine the elements of $\cQ$. However, as we now know, one
can also form
 $SL(2|2)$ invariants using the $\cE$ tensor. The simplest one requires at
least two $N$s and so occurs at five points. At first sight it
might seem that such objects are not actually defined on the
quotient space because they are only invariant under $SL(2|2)$ and
not under $GL(2|2)$, but this can easily be remedied. If we define

 \be
W:=\sdet\, u_{12} \,\sdet\,M_3 \ee

then $W$ transforms by

 \be
W\rightarrow (\sdet \,h_1)^2\, \sdet\, g\, W
 \ee

Now any $SL(2|2)$ invariant function of the $\{N_i\}$, not
invariant under $GL(2|2)$, will transform by some power $p$ of the
superdeterminant of $h_1$, and this can be compensated for by
multiplying it by $W^{-\ft{p}{2}}$. The resulting object will be
defined on the coset and is invariant under $PSL(4|4)$ but will
clearly not be invariant under $PGL(4|4)$. These are the
superconformal invariants which are not invariant under the
$U(1)_Y$ bonus symmetry.

It is easy to compare this formalism with the coordinate approach.
One can fix the local $GL(2|2)$ symmetries by choosing
representatives of the form

 \be
 X\mapsto s(X)=(1,\ X)\in\bbU
 \ee

at each point. One can then convert the invariants from the
Grassmannian formalism to those of the coordinate approach.

\subsection{Application}

We are now in a position to solve the Ward Identities for
$n$-point correlation functions of arbitrary operators in the
$N=4$ superconformal field theory in terms of propagators, known
functions of the coordinates and arbitrary functions of the
invariants which we have just seen how to construct. The
propagator factor absorbs the dilation weights of the operators;
in general, it is not uniquely determined, but any two admissible
propagator factors will be related to one another by an invariant
function. As in the case of the three-point function, each point
$j,j\neq 1$, can be translated to point 1, say, by using
$X_{1j}^{-1}$ in the appropriate representation. One then absorbs
the free indices by tensors $t$ which are monomials in the
variables $X_{12k}= X_{12}X^{-1}_{2k} X_{k1},\ k=3,\ldots n$ and
their inverses. The $t$-tensors may also involve $\d$s and $\cE$s,
and the presence of the latter requires the propagator factor to
be adjusted appropriately. In general, there will be many possible
$t$s and each independent one can be multiplied by an arbitrary
function of the $n$-point invariants and the coupling constant.
Any such solution will then be subject to the constraint that it
should be analytic in the internal $y$ coordinates. This
prescription solves the superconformal Ward Identities for all
components of all correlators of arbitrary gauge-invariant
operators in the theory. The schematic form of the solution is

 \be
<12..n> =\Pi_{j=2}^n \cR_j(X_{1j}^{-1})\cR'_j(X_{1j}^{-1})\sum_t
t^{\cR_2\ldots \cR_n;\cR'_2\ldots \cR'_n}_{\cR_1\cR'_1} P_t F_t
\ee

where the sum is over the possible tensors $t$, $P_t$ denotes an
appropriate propagator factor, which will depend on $t$ if the
latter involves factors of $\cE$, while $\cR_i$ and $\cR'_i,
i=1\ldots n,$ are the representations of the primed and unprimed
$\ggl(2|2)$ algebras under which the operators transform. Each
$F_t$ is an arbitrary function of the invariants and the coupling
which is restricted by analyticity. The formula makes sense if the
correlator includes long operators because the objects $\cR(X)$
can be defined for quasi-tensors \cite{Heslop:2001gp}. Moreover,
non-integral powers on the internal coordinates are cancelled by
similar factors in the propagators. The same comment does not
apply to the spacetime coordinates so that the formula is
perfectly compatible with analyticity in the internal bosonic
coordinates and anomalous spacetime dimensions.

The analysis of the restrictions imposed by demanding analyticity
was carried out in detail for the case of a four-point function in
$N=2$ with four operators of charge two in \cite{Eden:2000qp}
(this case arises in the $N=2$ decomposition of the four-point
function of $N=4$ supercurrents, for example). This type of
analysis can be expected to be quite complicated for a general
correlator. However, it was found that the OPE can help
significantly in the analysis of the analyticity properties of
four-point functions of one-half BPS operators in $N=4$ and  one
would expect that this should also be the case for more
complicated correlators \cite{Heslop:2002hp}. Other related
studies of four-point correlators of one-half BPS operators using
the OPE can be found in the literature
\cite{Arutyunov:2001mh,Dolan:2000ut,Arutyunov:2002fh,Arutyunov:2003ad}.

\section{Conclusions}

In this paper we have shown that the analytic superspace formalism
can be used fruitfully in the study of arbitrary gauge-invariant
operators in $N=4$ SYM. It is a convenient formalism to use for
the classification of explicit gauge-invariant operators,
simplifies the discussion of reducibility at threshold and allows
one to solve for correlation functions in terms of propagators,
explicit functions of the coordinates involving the $\ggl(2|2)$
representation matrices, and functions of invariants. We have also
shown how to obtain all superinvariants in analytic superspace. In
order to complete the analysis of all correlation functions in the
theory from the point of view of superconformal invariance one
needs to take into account the restrictions imposed by analyticity
in the internal even coordinates as we have just mentioned.

As has been shown elsewhere, the formalism can be used to give a
compelling non-perturbative argument in support of the
non-renormalisation of two- and three-point functions of arbitrary
protected operators, including one-quarter BPS series C operators,
one-eighth BPS series B operators and short series A operators.
However, as we have seen, the situation is more subtle for the
case of semi-protected operators which can arise in series A or B.
These operators have non-renormalised two-point functions even
though the three-point function with an extra supercurrent is not
$U(1)_Y$ invariant. This is because this correlator can have
$U(1)_Y$ charge at most equal to one. Generically, however, we
expect that three-point functions involving such an operator
together with two protected operators will be subject to
renormalisation effects.

The analysis of the properties of two- and three-point functions
is greatly facilitated by the use of the $\cE$ tensor. This makes
it easy to keep track of the $U(1)_Y$ properties of operators and
correlators and is also very useful in the construction of
nilpotent superconformal invariants.

\section*{Acknowledgements}

We thank  Burkhard Eden for many detailed discussions on invariants in 
analytic superspace.

This work was supported in part through European grants 
HPRN-CT-2000-00148 and HPRN-2000-00122  and  PPARC
grant PPA/G/O/2000/00451.

\vfill\eject

\section*{Appendix: from analytic to harmonic superspace}\la{harm}

Representations of the superconformal group are perhaps more
familiar as superfields on Minkowski superspace or harmonic
superspace. In this section we show how one can go from an
analytic superfield (generally with superindices) to a harmonic
superfield.

Let us suppose we have two coset spaces $M_i=H_i\bsh G, i=1,2$
such that $H_1\subset H_2$. Then $M_1$ is a fibre bundle over
$M_2$ with fibre $Y=H_1\bsh H_2$. A tensor field $f_2$ on $M_2$,
transforming under an induced representation of $G$, is equivalent
to an equivariant field $F:G\rightarrow V$, by which we mean
$F(h_2 u)=\cR(h_2) F(u)$, where $u\in G,\ h_2\in H_2$ and where
$\cR$ denotes a linear representation of $H_2$ on the vector space
$V$. If $s_2(x_2)$ denotes a local section $M_2\mapsto G$, then
the tensor field is given locally by $f_2(x_2)=F(s(x_2))$. We can
define a related field $f_1(x_1)$ on $M_1$ as follows: in local
coordinates we can write $x_1=(x_2,y)$, $y$ being the fibre
coordinates, and we can write a local section $s_1:M_1\rightarrow
G$ as $s_1(x_1)=s(y)s_2(x_2)$ where $s(y)$ is a section
$Y\rightarrow H_2$. We can then put
$f_1(x_1)=F(s(x_1))=F(s(y)s_2(x_2))=\cR(s(y))f_2(x_2)$. This
construction gives a field on $M_1$ which transforms  as follows
under $G$:

 \be
 f_1(x_1)\mapsto (g\cdot f_1)(x_1)=\cR(h_1(x_1,g))f_1(x_1\cdot g)
 \ee

where

 \be
 s(x_1)g=h_1(x_1,g)s(x_1\cdot g)
 \ee

As a tensor under $H_1$ $f_1$ will in general transform reducibly,
but in the case of interest it is easy to select the required
irreducible representation.

We shall now show how to construct fields on $(4,2,2)$ harmonic
superspace starting from fields on $(4,2,2)$ analytic superspace.
The relevant fibration is

\be
 \begin{picture}(192,60)(0,0) \put(0,0){\makebox[0pt][l]{$\bt
 \hspace {2em}
 \ominus \hspace{2em}\bt\hspace{2em}\times\hspace{2em}\bt
 \hspace{2em}\ominus \hspace{2em}\bt$}
 \rule[.5ex]{17.1em}{.1ex} }
\put(96,25){$\downarrow$} \put(0,50){\makebox[0pt][l]{$\bt
 \hspace {2em}
 \otimes \hspace{2em}\bt\hspace{2em}\times\hspace{2em}\bt
 \hspace{2em}\otimes \hspace{2em}\bt$}
 \rule[.5ex]{17.1em}{.1ex} }
 \end{picture} 
 \ee

where the top line represents $(4,2,2)$ harmonic superspace. This
diagram indicates that the fibres in this case factorise into
left-hand and right-hand parts. These can both be represented by
the subdiagram

\be
 \begin{picture}(30,50)(0,-40) \put(0,0){\makebox[0pt][l]{$\bt
\hspace {2em}
 \otimes \hspace{2em}\bt$} \rule[.5ex]{5.5em}{.1ex} }
\put(32 ,-20){$\downarrow$}
 \put(0,-40){\makebox[0pt][l]{$\bt
\hspace {2em}
 \ominus \hspace{2em}\bt$} \rule[.5ex]{5.5em}{.1ex} }
\end{picture}\la{sfields}
\ee

The top line here represents the coset space $ H\bsh GL(2|2)$
where $H$ is the set of matrices of the form

\be \left( \ba{cc|cc} \bt & \bt & 0 &0\\
                     \bt & \bt & 0 &0 \\ \hline
                      \bt & \bt &\bt &\bt\\
                     \bt & \bt &\bt &\bt
\ea \right) \ee

where the bullets represent non-zero elements. The bottom line has
no crosses and thus corresponds to a point. So in this context one
takes a linear representation of $GL(2|2)$ and lifts it to the
coset. This coset space has only fermionic coordinates so that we
are effectively rewriting tensors (or quasi-tensors) as
 ``superfields'' which depend on the odd
coordinates of the coset but not on the space-time coordinates. We
can do this with both factors, take the product and allow the
fields so obtained to depend on the coordinates of analytic
superspace. In this way we shall have written the latter as fields
on harmonic superspace.

\subsection*{Representations of supergroups as superfields.}

We shall  write an arbitrary finite dimensional irreducible
representation of $GL(2|2)$ as a superfield on the coset space $
H\bsh GL(2|2)$

A coset representative of this space is given by

\be s(\r)^B{}_A=\left( \ba{c|c} \d^b{}_a& \r^{\b}{}_a\\ \hline
0&\d^{\b}{}_{\a} \ea \right). \ee

The Dynkin diagram for this space is

\be
 \begin{picture}(30,10)(0,0) \put(0,0){\makebox[0pt][l]{$\bt
\hspace {2em}
 \otimes \hspace{2em}\bt$} \rule[.5ex]{5.5em}{.1ex} }
 \end{picture}
\ee

and we see that the cross splits the diagram into two $GL(2)$
representations. We accordingly split the indices $A$ as
$A=(\a,a)$ with $a,\a$ both running from 1 to 2. We shall write a
tensor carrying $GL(2|2)$ indices $A$ as a superfield carrying
$GL(2) \xz GL(2)$ indices $(a,\a)$.  A representation of $GL(2|2)$
will decompose into different representations under $GL(2) \xz
GL(2)$ but we need to select the irreducible one. From the
triangular structure of the isotropy group one sees that this
representation will be the one with the maximum number of $a$-type
indices. A simple example is given by the defining representation
$V_A$. The required field is

\be
 v_a:= V_B (s^{-1})^B{}_a = V_a - V_{\b}\r^{\b}_a
\ee

which clearly contains both $V_a$ and $V_\a$. On the other hand
$V_B (s^{-1})^B{}_\a=V_\a$ and is in fact not irreducible under
$H$. (We use the inverse of $S$ in order to get a left
representation.) In the general case the index structure of the
required superfield can be read off from the Young
tableau~\eq{YT}:

\be \setlength{\unitlength}{.2mm}
\begin{picture}(440,120)(60,40)
\put(100,140){\framebox(100,20){\tiny $m_2$}}
\put(100,120){\framebox(60,20){\tiny $m_1$}}
\put(60,40){\framebox(20,120){\tiny $m_4$}}
 \put(80,80){\framebox(20,80){\tiny $m_3$}}
\put(200,120){$\rightarrow$} \put(240,140){\framebox(100,20){\tiny
$m_2$}} \put(240,120){\framebox(60,20){\tiny $m_1$}}
\put(340,120){$\xz$} \put(380,140){\framebox(100,20){\tiny $m_4$}}
\put(380,120){\framebox(60,20){\tiny $m_3$}} \put(120,40){$GL(2|2)
\quad \supset \quad GL(2) \quad \xz \quad GL(2)$}
\end{picture}\la{YT}
 \ee

so the superfield has $m_1 +m_2$ $\a$ indices and $m_3+m_4$ $a$
indices arranged according to these Young tableau. One converts
from a tensor to a superfield by multiplying the tensor by $m_3 +
m_4$ copies of $s^A{}_b$ and $m_1+m_2$ copies of $s^A{}_{\b}$,
contracting the tensor indices with the corresponding upstairs
indices of $s^A{}_B$, and putting the remaining indices into the
correct representation.

Note that the superfield can be straightforwardly converted into a
superfield with $m_2-m_1$ $\a$ indices and $m_4-m_3$ $a$ indices
by applying $\e$-tensors. This new superfield will now transform
with a $\com$-charge.

As another example consider the symmetric representation $V_{AB}$
with Young tableau $<0,0,1,1>$. It can be written as a
``superfield'' $v_{ab}$ where

\be v_{ab}=V_{AB} (s^{-1})^B{}_b (s^{-1})^A{}_a (-1)^{(A+a)b}=
V_{ab} - 2V_{\b}{}_{[a}\r^{\b}{}_{b]} + V_{\a \b} \r^{\a}_a
\r^{\b}_{b} \ee

The minus sign can be viewed as follows: we want to contract
$(s^{-1})^A{}_a$ directly next to the $A$ index `before' the $B$
index. Since we can not do this we commute $s$ past the $b$
picking up the minus sign in the process. In this example the
$a,b$ indices are anti-symmetric and so one can multiply by $\e$
to obtain $v:=\e^{ab} v_{ab}$.

Notice that in this example, as for the fundamental
representation, the superfields do not have a maximum possible
theta expansion, and are thus `short'. This shortness can be
expressed as a constraint on the superfield, which in the above
two examples reads

\be
 D_{\a(a} v_{b)}=0 \qquad
 D_{\a}^{a}D^{\a b} v=0 \la{constraints}
\ee

where $D^a_{\a}:=\del/\del \r^{\a}_a$, and where indices of both
types are raised or lowered with epsilon tenors. These are the
origins of the quarter BPS constraints and the constraints on
short scalar fields which saturate the series A unitary bounds.

In~\cite{Heslop:2002hp} the representations of $GL(2|2)$ were
classified into long and short representations, the long ones
corresponding to $m_3>1$ and the short ones corresponding to
$m_3=1$ or $m_3=0$ (with the canonical choice $m_1=0$). On writing
the tensor representation as a superfield one finds that the long
or typical representations have full superfield expansions (in
$\r$) whereas the short or atypical representations (such as the
ones given in this example) have some terms missing in the theta
expansion. For example one can easily convince oneself that the
representation with Young tableau $<0,0,2,2>$ has a full $\r$
expansion and hence is a long representation.

\subsection*{Analytic superfields as harmonic superfields.}

We can now apply this construction in order to write superfields
on $(4,2,2)$ analytic superspace as superfields on $(4,2,2)$
harmonic superspace. We can simply apply the results of the
previous subsection to the left- and right-handed $GL(2|2)$
isotropy groups.

We have the coset representatives

\be s(\r)^A{}_B=\left( \ba{c|c} \d^{\a}{}_{\b} & \r^{\a}{}_b\\
\hline 0&\d^a{}_b\ea \right) \qquad s'(\h)_{A'}{}^{B'}=\left(
\ba{c|c} \d_{\a'}{}^{\b'}& 0\\ \hline
\h_{\a'}{}^{b'}&\d_{a'}{}^{b'}\ea \right). \ee

$\r$ and $\h$ will become the extra coordinates that harmonic
superspace has in addition to those of analytic superspace.  To
obtain a superfield on harmonic superspace from one on analytic
superspace, simply multiply the analytic superfield on the right
by $s^{-1}(\r)$ and on the left by $s'(\h)$ in the way described
in the previous section and then choose the component with the
maximum of internal indices.

For example, a one-half BPS state $A$ has no indices, so it lifts
trivially to harmonic superspace. It does not depend on the extra
$(\r,\h)$ coordinates as we would expect. The independence of $A$
of $(\r,\h)$ can be expressed in terms of differential
constraints:

\be D_{\a}^aA=\bar D_{\a' }^{a'}A=0. \ee

where, in the coordinates $(x,\l,\p,y,\r,\h)$, we have defined
$D_{\a}^a :=\del/\del \r^{\a}_a ; \ \
   \bar D_{\a'}^{ a'} :=\del/\del  \h^{\a'}_{a'}$.

Now consider a one-quarter BPS operator with Dynkin labels of the
form $[001d100]$ where $Q=d+2$ is the $Q$ charge. For any value of
$d$ an operator of this type is represented on analytic superspace
by a covector field, $V_{A'A}(x,\l,\p,y)$, say.  It lifts to a
superfield $v_{a'a}(x,\l,\p,y,\r,\h)$ in harmonic superspace where

 \bea
 v_{a'a} &=& s'_{a'}{}^{B'}V_{B'B}(s^{-1})^B{}_a\nn\\
 & =&V_{a'a} -V_{a' \b} \r^{\b}{}_a + \h_{a'}{}^{\b'}V_{\b' a} -
\h_{a'}{}^{\b'}V_{\b' \b}\r^{\b}{}_a .
 \eea

This is also a constrained superfield on harmonic superspace as
can be seen by the fact that it has a short $\r$ expansion.
Equation~\eq{constraints} shows that the constraints are given by

\be D_{\a(a} v_{b)b'}=\bar D_{\a'(a'} v_{bb')} =0. \ee

This example makes it clear that we can write any analytic
superfield as a harmonic superfield in terms of the coordinates
$(x^{\a \a'},\l^{\a a'},\p^{a \a'},y^{a a'},\r^{\a}_a,\h^{a'
  \a'})$.
Whilst analytic superfields are always unconstrained we see that
on lifting to harmonic superspace the resulting harmonic
superfields often satisfy constraints. The constraints satisfied
by irreducible superconformal representations on (a slightly
different) harmonic superspace are given in~\cite{Ferrara:2000eb}.

Harmonic superspace is Minkowski superspace extended by the
internal coordinates $y$ and therefore has the standard
coordinates of complex super Minkowski space $(x^{\a \a'}_M,
\th^{\a i}, \vf^{\a'}_i)$  together with the internal coordinates
$y^{aa'}$. If we split the odd Minkowski coordinates in two pairs,
$\th^{\a i}=(\th^{\a}_a,\th^{\a a'})$ and $\vf^{\adt}_i=(\vf^{\a'
a}\vf^{\adt}_{a'})$, then the two sets of coordinates for harmonic
superspace are related as follows:

\be \ba{rcl} x^{\a \a'}&=& x_M^{\a \a'} +{1\over2}\left(
\th^{\a_a} \vf^{\a' a} -
\th^{\a a'}\vf^{\a'}_{a'}+2y^{aa'}\th^{\a}_a \vf^{\a'}_{a'}\right)\\
\l^{\a a'}&=&\th^{\a a'}- \th^{\a}_a y_{aa'}\\
\p^{a \a'}&=&\vf^{\a' a} + y^{aa'} \vf^{\a'}_{a'}\\
\r^{\a}_a&=&\th^{\a}_a\\
\h_a^{\a' }&=&\vf_a^{\a' }\ea\la{coords} \ee

We can rewrite this in the language of GIKOS by introducing a
matrix $u_I{}^i\in SL(4)$. For $(4,2,2)$ harmonic superspace we
split the internal coset index into two doublets $I=(r,r')$ and
convert $SL(4)$ indices to coset indices using $u$ and its inverse
in the standard way. In this notation G-analytic fields depend
only on $x^{\a\a'},\th^{\a r'}$ and $\vf^{\a'}_r$, where

 \be
 x^{\a\a'}=x_M^{\a\a'}+{1\over 2}(\th^{\a r}\vf^{\a'}_r - \th^{\a
 r'}\vf^{\a'}_{r'})
 \ee

The link between the two notations is made by choosing a gauge for
$u$ of the form

 \be
 u_I{}^i=\left(\ba{ll} \d^a{}_b & y^{a b'}\nn\\
 0 & \d_{a'}{}^{b'}\ea\right)
 \ee

Finally, to go to real superspace in the conventions of
\cite{Howe:md,Hartwell:1994rp} one needs to replace each Minkowski
superspace coordinate $z$ by $-iz$ and then impose the reality
conditions, $x=\bar x,\vf=\bar\th$. In the real case the internal
symmetry group becomes $SU(4)$. Note also that the $x$ coordinate
we are using is the usual one for analytic space which is often
denoted $x_A$ in the literature.

\end{document}